\documentclass[aps,prl,twocolumn, epsfig]{revtex4}
\usepackage[dvips]{graphics}
\usepackage{graphicx}
\usepackage{amsfonts}
\usepackage{amssymb}
\usepackage{amsmath}
\begin{document}

\title {Matter wave quantum dots (anti-dots) in ultracold atomic Bose-Fermi mixtures}
\author{Mario Salerno}
\affiliation{Dipartimento di Fisica "E.R. Caianiello",
 Universit\'a di Salerno, via S. Allende, I-84081 Baronissi (SA),\\
         Consorzio Interuniversitario per le Scienze Fisiche della
         Materia (CNISM), Unita' di Salerno,
         \\ Istituto Nazionale
         di Fisica Nucleare (INFN), Gruppo Collegato di Salerno, Italy.}
\begin{abstract}
The properties of ultracold atomic Bose-Fermi mixtures in external
potentials  are investigated and the existence of gap solitons of
Bose-Fermi mixtures in optical lattices demonstrated. Using a
self-consistent approach we compute the energy spectrum and show
that gap solitons can be viewed as matter wave realizations of
quantum dots (anti-dots) with  the bosonic density playing  the
role of trapping (expulsive) potential for the fermions. The
fermionic states trapped in the condensate are shown to be at the
bottom of the Fermi sea and therefore well protected from thermal
decoherence. Energy levels, filling factors and parameters
dependence of gap soliton quantum dots are also calculated both
numerically and analytically.
\end{abstract}
\maketitle

\section{Introduction}

The possibility to cool atomic Fermi gases below the Fermi
temperature \cite{DeMarco} with the developement of the
sympathetic cooling technique \cite{Hulet}  has  raised a
significant amount of interest in theoretical and experimental
studies of atomic Fermi gases embedded in Bose-Einstein
condensates (BEC). Such systems have very interesting features
which allow to explore a whole class of phenomena, ranging from
strongly correlated effects in fermionic atomic gases to novel
superfluid phases in Bose-Fermi mixtures (BFM) analogous to BCS
superconductivity. The possibility to trap fermionic states into
localized bosonic components of BFM makes these systems also of
potential interest for quantum computing. The basic element of a
quantum computer is the so called quantum bit, a special quantum
state usually realized with a two-level system, which is
manipulated by logical quantum gates (unitary operators) and which
must be preserved intact for an extended period of time to allow
computation \cite{Nielsen}. An important role in this context is
played by the interaction of the quantum system with the
environment (thermal reservoir) which unavoidably produces
decoherence. For an ideal q-bit one would like to keep the quantum
system  well protected from decoherence but not completely
isolated from the environment to allow gate manipulations. This
condition could be realized in Bose-Fermi mixtures by trapping
fermionic states into localized BEC, so to form quantum dots with
two (or more) fermionic levels inside, and by using Feshbach
resonances to control the boson-fermion interaction, i.e. the
coupling of the quantum bit with the environment.

The aim of this paper is twofold. From one side, we investigate
the possibility of using ultracold Bose-Fermi mixtures to form
matter wave quantum dots with two or more fermionic levels inside.
To this regard we consider solitonic BEC states as trapping
(expulsive) dot (antidot) potentials and show that the fermionic
bound states formed inside are at the bottom of the Fermi sea and
therefore well protected by thermal decoherence.

From the other side, we investigate the spectral properties of a
BFM both in parabolic and in periodic potentials and demonstrate
the existence of BFM gap-solitons in optical lattices (OL).

To this regard, we derive equations of motion for the mixture
adopting a mean field approximation for the bosons and assuming
non interacting (spin polarized) fermions in true quantum regime.
These equations are solved by means of a self-consistent procedure
and the spectral properties of BFM, both in parabolic traps and in
optical lattices, derived. In particular, we show that fermionic
bound state levels at the bottom of the Fermi sea can be trapped
inside gap-solitons for both repulsive and attractive boson-boson
interactions. We also show that, as the boson-fermion interaction
is varied, the trapped levels undergo avoid crossings or level
crossings according to the symmetry of the corresponding
wavefunctions with respect to the OL. Analytical expressions for
energy levels and for the number of fermionic bound states trapped
inside attractive gap solitons are also derived and the results
compared with those obtained from direct numerical self-consistent
calculations. We find that the filling factor of a gap soliton
follows the law $n_f\propto \sqrt{|\chi_{bf}|N_b}$, where $N_b$ is
the number of bosons, $n_f$ the number of fermions trapped in the
condensate and $\chi_{bf}$ the strength of the boson-fermion
interaction. The possibility to change $\chi_{bf}$ in a wide
interval of values using Feshbach resonances suggests the
possibility of an experimental check of this results. In
particular we expect that, as the boson-fermion interaction is
increased, the progressive filling of the gap soliton should give
rise to a depletion of fermions in the OL which follows the above
law and which could be monitored by imaging techniques.

The paper is organized as follow. In section II we introduce the
model and derive quasi-classical equation of motion of Bose-Fermi
mixtures directly from a second quantized formulation of problem.
In section III we investigate the energy spectrum for BFM in
attractive and repulsive parabolic potentials and discuss quantum
dots (antidots) analogies. In section IV we consider BFM in
optical lattices and show the existence of gap solitons both for
repulsive and attractive boson-boson interactions. The number of
fermions  trapped in a BFM gap soliton and the energies of the
bound states are also calculated both numerically and
analytically. Finally, in the last section the main results of the
paper are briefly discussed and summarized.

\section{Equations of motion for a Bose-Fermi
mixture} To derive equations of motion of a BFM it is convenient
to start from the  second quantized  Hamiltonian of a set of
bosons and fermions in interaction
\begin{eqnarray}
&&\hat H =\int dr \hat\psi_b^\dagger(r)\hat H_b^0 \hat\psi_b
(r)+\int dr \hat\psi_f^\dagger(r)\hat H_f^0 \hat\psi_f (r)+
\nonumber \\ && \frac12 \int\int dr
dr'\hat\psi_b^\dagger(r)\hat\psi_b^\dagger(r')U_{bb}(r-r')
\hat\psi_b(r')\hat\psi_b(r)+ \nonumber \\&& \int\int dr
dr'\hat\psi_b^\dagger(r)\hat\psi_f^\dagger(r')U_{bf}(r-r')
\hat\psi_b(r')\hat\psi_f(r),
\end{eqnarray}
where $\hat H_b^0 = -\frac{\hbar^2}{2 m_b}\nabla^2+ V_{ext}^b,
\;\hat H_f^0=-\frac{\hbar^2}{2 m_f}\nabla^2+ V_{ext}^f$, are
single particle hamiltonians accounting for the kinetic energy and
for the external potential confining the bosons and the fermions,
respectively, while $\hat\psi_b^\dagger(r), \hat\psi_b(r)$,
$\hat\psi_f^\dagger(r), \hat\psi_f(r)$ are usual bosonic and
fermionic field operators satisfying
$[\hat\psi_b(r),\hat\psi_b(r')]=
[\hat\psi_b^\dagger(r),\hat\psi_b^\dagger(r')]=0,
\;[\hat\psi_b(r),\hat\psi_b^\dagger(r')]=\delta(r-r')$ (and
similar relations for the fermionic fields with commutator
replaced by the anti-commutator). Notice that the fermions are
assumed spin polarized and therefore non interacting, so that only
the boson-boson and boson-fermion interactions appear in the
Hamiltonian. Using the above operator algebra, the Heisenberg
equation of motion for the bosonic and fermionic fields can be
written as
\begin{eqnarray}
i\hbar\frac{\partial \hat\psi_b}{\partial t}= \hat H_b^0
\hat\psi_b(r)+ \int dr'
\hat\psi_b^\dagger(r')U_{bb}(r-r')\hat\psi_b(r')\hat\psi_b(r)
\nonumber
\\+ \int
dr'\hat\psi_f^\dagger(r')U_{bf}(r-r')\hat\psi_f(r')\hat\psi_b(r),
\nonumber \\
i\hbar\frac{\partial \hat\psi_f}{\partial t}= \hat H_f^0
\hat\psi_f(r)+ \int dr'
\hat\psi_b^\dagger(r')U_{bf}(r'-r)\hat\psi_f(r')\hat\psi_f(r).\nonumber
\end{eqnarray}
For dilute mixtures one can replace the interaction potentials
with the effective interactions
\begin{equation}
U_{bb}(r-r')=g_{b} \delta(r-r'),\;\; U_{bf}(r-r')=g_{bf}
\delta(r-r'),
\end{equation}
where the coupling constants $g_{b}, g_{bf}$, are related to the
s-wave scattering lengths $a_{b}, a_{bf}$, by the relations:
$g_{b}= 4\pi \hbar^2 a_{b}/m_b$,  $g_{bf}=2\pi \hbar^2
a_{bf}/m_{bf}$,  where $m_{bf}$ denotes the reduced mass
$m_{bf}=m_b m_f/ (m_b+m_f)$. Using these potentials, the above
equations  reduce to
\begin{eqnarray} && i \hbar \frac{\partial \hat\psi_b(r,t)}{\partial t}=
\left(\hat H_b^0 +g_{b} \hat\rho_b(r,t) + g_{bf}
\hat\rho_f(r,t)\right) \hat\psi_b(r), \nonumber
\\&& i \hbar\frac{\partial \hat\psi_f(r,t)}{\partial t} =\left( \hat
H_f^0 + g_{bf} \hat\rho_b(r,t)\right) \hat\psi_f(r),
\label{heiseqn2}
\end{eqnarray}
where $\hat\rho_b=\hat\psi_b^\dagger\hat\psi_b$ and
$\hat\rho_f=\hat\psi_f^\dagger \hat\psi_f$, are the bosonic and
fermionic  particle densities, respectively. Expanding Bose and
Fermi fields in terms of the usual creation and annihilation
operators $\hat a^\dagger, \hat a$ and $\hat c^\dagger, \hat c$,
for bosons and fermions, respectively,
$$
\hat \psi_b(r,t)= \sum \psi^b_n \hat a_n, \;\;\; \hat \psi_f(r,t)=
\sum \psi^f_n \hat c_n,
$$
and replacing  all field operators appearing in the bosonic
equation with their expectation values on the ground state (mean
field approximation)
\begin{eqnarray}
\hat\psi_b \rightarrow \langle \hat\psi_b \rangle\approx
\psi^b_0(r,t),\;\; \hat \rho_b \rightarrow \langle \hat \rho_b
\rangle \approx N_b |\psi^b_0(r,t)|^2, \nonumber\\
\hat \rho_f \rightarrow \langle \hat \rho_f \rangle \approx
\sum_{n=1}^{N_f} |\psi^f_n(r,t)|^2,\nonumber
\end{eqnarray}
we have that Eqs. (\ref{heiseqn2}) reduce to
\begin{eqnarray} && i \hbar \frac{\partial \psi^b_0}{\partial t}=\left[\hat
H_b^0+g_{b} N_b |\psi^b_0|^2 + g_{bf}\left( \sum_{n=1}^{N_f}
|\psi^f_n|^2 \right)\right] \psi^b_0, \nonumber
\\&& \sum_n \left[ i \hbar\frac{\partial \psi^f_n}{\partial t} -\hat
H_f^0 \psi^f_n- g_{bf} N_b |\psi^b_0|^2 \psi^f_n\right] \hat
c_n=0.\label{eq2}
\end{eqnarray}
The  completeness of the fermionic Fock space implies that the
second equation in (\ref{eq2}) can be satisfied only if the terms
in the square bracket vanishes for all $n$ , this giving the
following $N_f+1$ coupled equations
\begin{eqnarray} && i \hbar \psi^b_t=
\left[-\frac{\hbar^2}{2 m_b}\nabla^2+V_{ext}^b+
g_{b}N_b|\psi^b|^2+g_{bf} \rho_{f} \right]\psi^b, \nonumber
\\&& i \hbar \psi^f_{n_t}=
\left[ -\frac{\hbar^2}{2 m_f}\nabla^2+V_{ext}^f + g_{bf} N_b
|\psi^b|^2 \right]\psi^f_n,\label{bose-fermi-eqns}
\end{eqnarray}
with $\rho_f=\sum_{i=1}^{N_f} |\psi_i^f|^2$ the fermionic density
and $\psi^b_0\equiv \psi^b$.  From the above  derivation it is
clear that while bosons are treated in a classical mean field
approximation the spin polarized fermions are in true quantum
regime due to the lacking of interaction among them. Also note
that for $g_{bf}=0$ the above equations separate into the
Gross-Pitaevskii equation for the BEC part and $N_f$ identical
(due to the lack of interaction among fermions) Schr\"odinger
equations for the fermionic part of the mixture. Equations
(\ref{bose-fermi-eqns}) were also derived in
\cite{Karpiuk1,Karpiuk2} by means of a Lagrangian density
approach.

In the following we shall consider Eq. (\ref{bose-fermi-eqns}) in
a quasi one dimensional context by assuming the bosons and
fermions confined in a cylindrical trap with transverse  trapping
frequency $\omega_{\perp}$ and negligible x-axial confinement. We
also assume that the ground state energy of the transverse
trapping potential $\hbar \omega_{\perp}$ is larger than the 3D
ground state energies of the bosons and the fermions. Within this
approximation the bosonic and fermionic wavefunctions can be
factorized as $\psi=\psi(x)\psi_{\perp}(y,z)$ with the transverse
wavefunction taken as the ground state of the trap potential
independently of the longitudinal behavior and statistics
\cite{das}. By integrating over transverse coordinates one obtains
effective 1D equations which are formally identical to Eqs.
(\ref{bose-fermi-eqns}) but with boson-boson and boson-fermion
interactions rescaled according to: $g_{b}\rightarrow 2 \hbar
\omega_{\perp} g_{b},\;\;\; g_{bf}\rightarrow 2 \hbar
\omega_{\perp} g_{bf}$.

Notice that the fact that fermionic and bosonic densities appear
in the equations of bosons and fermions, respectively,  suggests
to consider them as effective potentials and solve the
quasi-classical equation of motion by means of a self consistent
procedure. Starting with a trial function for the bosonic part, we
solve the quantum eigenvalue problems for the fermionic part,
compute the fermionic density, derive the bosonic wavefunction by
solving the bosonic equation and  iterate the procedure until
convergence is reached (an efficient implementation of this scheme
for single BEC was discussed in \cite{salerno}). In the following
we apply this method for BFM in parabolic traps and in optical
lattices, concentrating, for simplicity, on the quasi one
dimensional case.

\section{Bose-Fermi mixtures in a parabolic trap} In this section we
consider BFM in a trap potential of the form $
 V_{trap}= \frac \varepsilon 2 m   \omega^2 (x-x_0)^2 $
and assume for simplicity $m_b=m_f$, $V_{ext}^b=
V_{ext}^f=V_{trap}$ (the extension to the generic case is
straightforward). We normalize Eqs. (\ref{bose-fermi-eqns}) by
measuring space in harmonic oscillator length units
$a_0=\sqrt{\hbar/m \omega}$, time in units of $\hbar/E_0$, with
$E_0$ the zero point energy of the oscillator. Introducing the
parameter $\chi_{b (bf)}$  as $g_{b (bf)}= (8 \pi \hbar^3
\omega_{\perp} a_{b_0 (bf_0)}/m) \chi_{b (bf)},\;$ where
$a_{{b}_0}, a_{{bf}_0}$ are scattering lengths in zero magnetic
field (we use $\chi_{b (bf)}$ to change nonlinearities by means of
external magnetic fields via Feshbach resonances), and rescaling
wavefunctions according to $\psi^b \rightarrow (4 \pi g_{{b}_0}
N_b /m \omega a^2_{\perp})^{1/2} \psi^b$, $\psi^f \rightarrow (4
\pi g_{{bf}_0}/m \omega a^2_{\perp}) \psi^f$, with
$a_\perp=(\hbar/m \omega_{\perp})^{1/2}$, we obtain the normalized
equations
\begin{eqnarray} && i \psi^b_t=
-\psi^b_{xx}+ V \psi^b + \chi_{b}|\psi^b|^2 \psi^b +\chi_{bf}
\rho_f \psi^b, \nonumber
\\&&
i \hbar {\psi^f_i}_t= -{\psi^f_i}_{xx}+ \left[ V + \alpha
\chi_{bf} |\psi^b|^2 \right]\psi^f_i, \label{norm-eqns}
\end{eqnarray}
where  $V= \varepsilon (x-x_0)^2$ and  $\alpha=a_{bf}/a_{b}$
assumed in the following of the order of unity (notice that with
this normalization the physical number of bosons $N_p$ is obtained
from the dimensionless number $N_b$ as $N_p=m\omega a_{\perp}^2
N_b/4 \pi g_{{b}_0}$).
\begin{figure}[htb]
\centerline{
\includegraphics[width=4.2cm,height=4.2cm,clip]{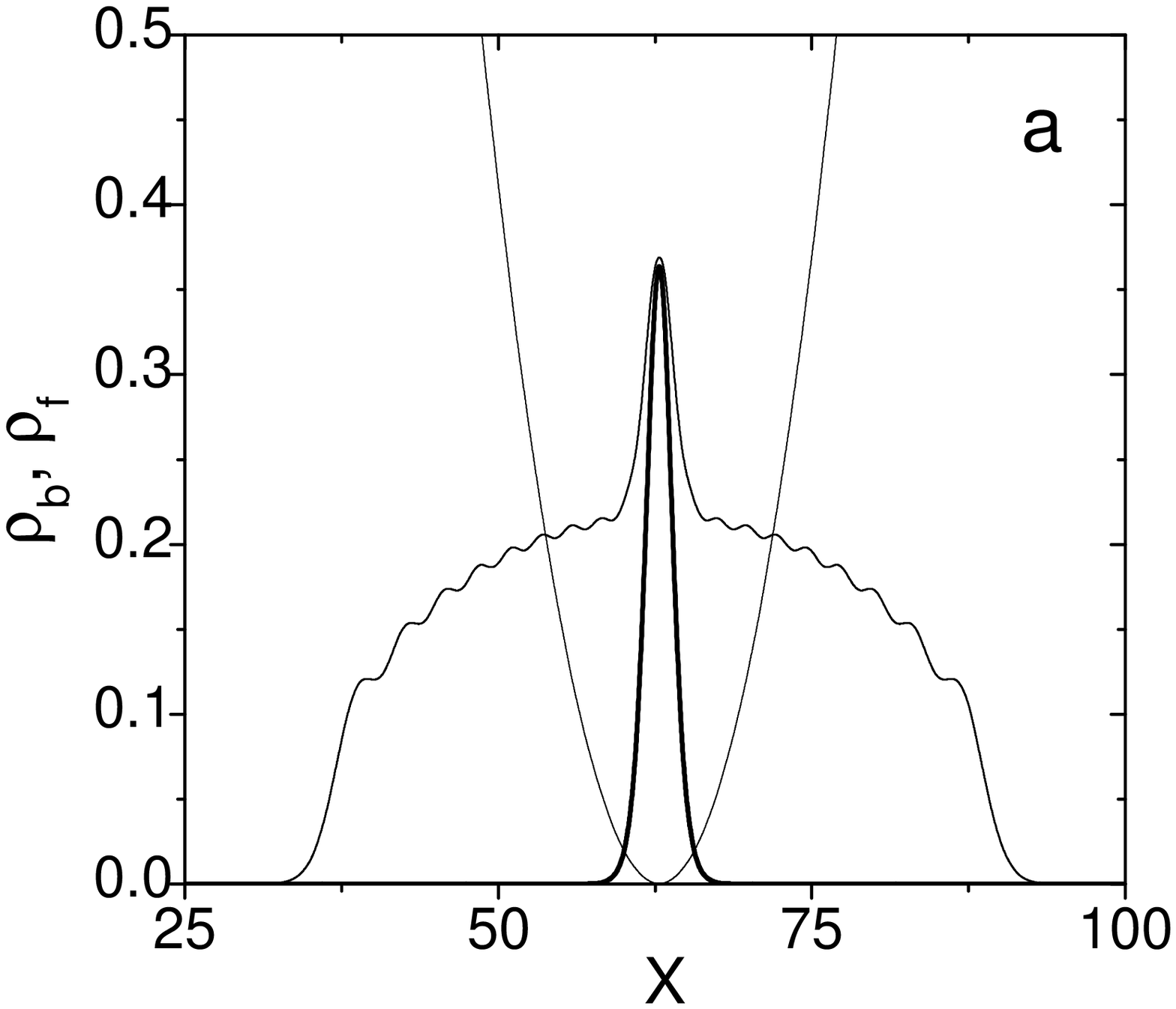}
\includegraphics[width=4.2cm,height=4.2cm,clip]{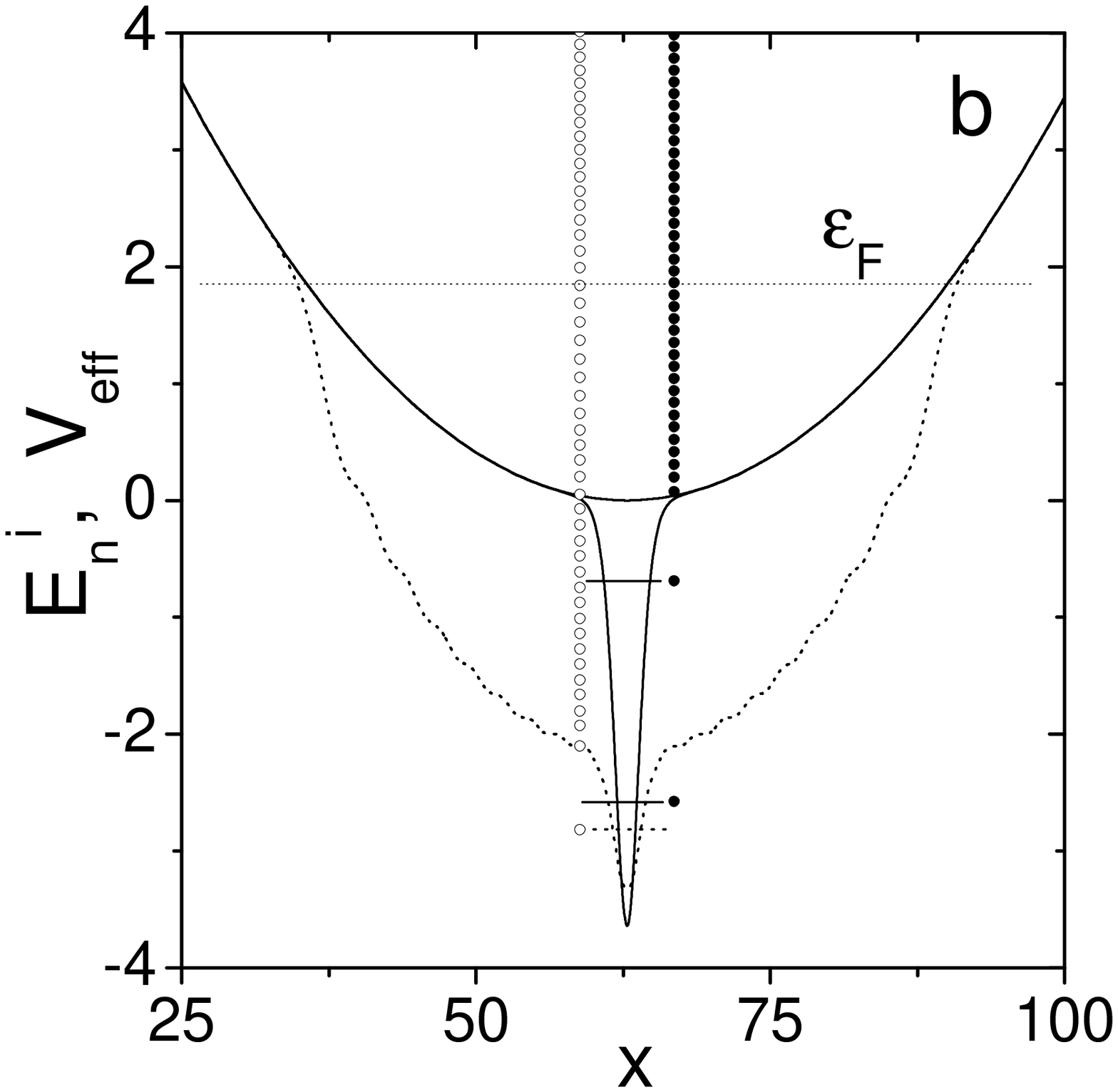}
} \centerline{
\includegraphics[width=4.2cm,height=4.2cm,clip]{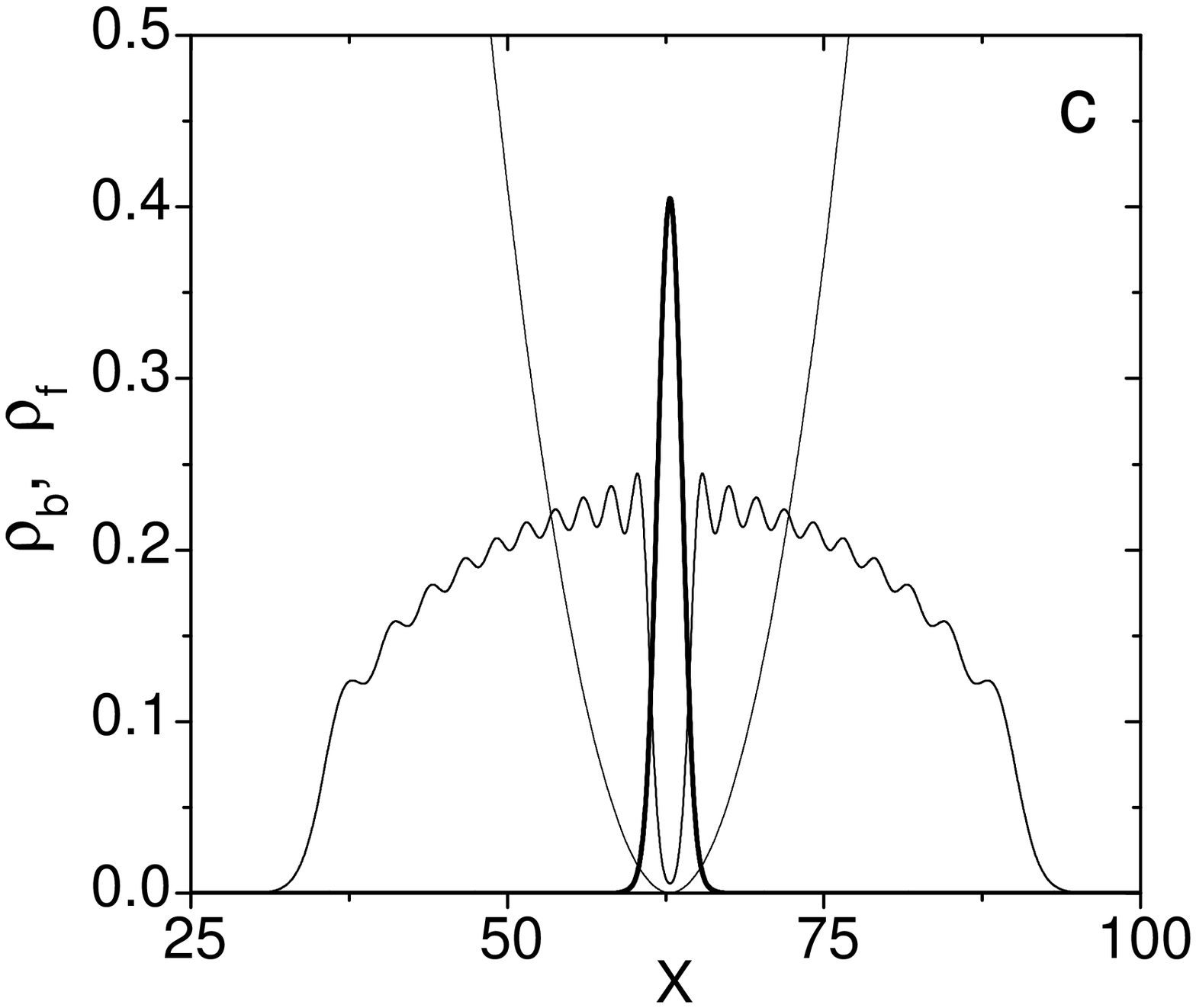}
\includegraphics[width=4.2cm,height=4.2cm,clip]{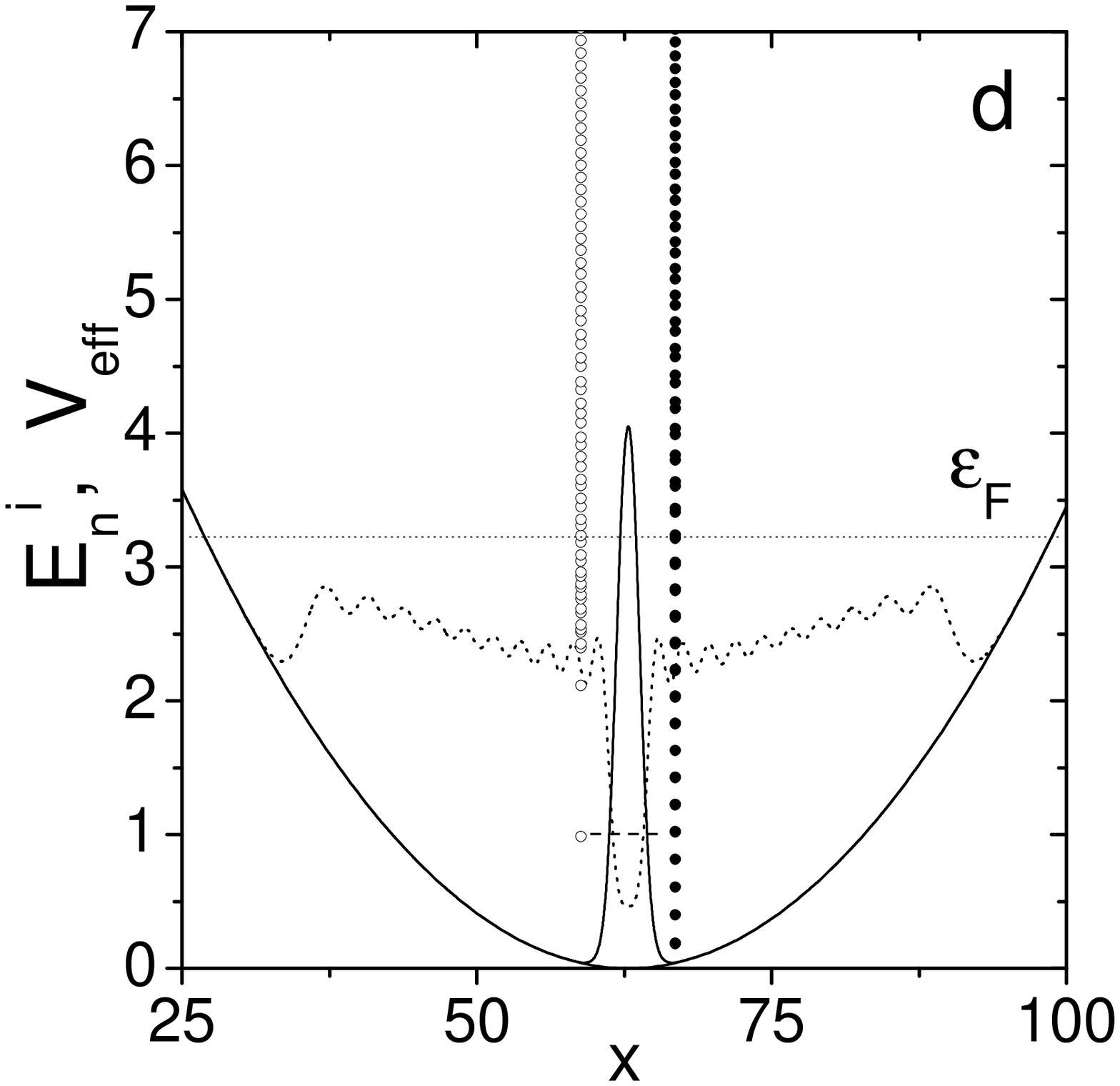}
} \caption{(a). Fermionic and bosonic densities for the case of
$\chi_{b}=1, \chi_{bf}=-10$, $N_f=20$ and normalized number of
bosons $N_b=1$, for a parabolic trap of strength
$\varepsilon=0.0025$ centered at $x_0=L/2=20\pi$. (b) Effective
potentials and energy levels of the Bose (open circles and dotted
line) Fermi (filled circles and continuous line) mixture with
densities given in panel(a) (horizontal line refers to the Fermi
level). Panels (c) and (d) are the same as panels (a),(b),
respectively, except for $\chi_{bf}=10$. Plotted quantities are in
normalized units.} \label{fig1}
\end{figure}

In Fig.\ref{fig1}a we depict the bosonic and fermionic densities
obtained with the self-consistent approach for a BFM with
repulsive bosonic interactions ($\chi_{b}>0$) and attractive
bosons fermion interaction ($\chi_{bf}<0$). We see that while the
bosons remain confined in the middle of the trap, the fermionic
density is quite extended in space and has a hump corresponding to
the bosonic component. This hump originates from the contribution
of the fermions trapped by the BEC acting on the fermions as a
potential well (trapping impurity). Similar results were also
reported in \cite{Karpiuk2}.

An interesting problem which has not been investigated is the
distribution of the bosonic and fermionic levels in their
effective potentials. This is done in Fig. \ref{fig1}b from which
we see the presence of two fermionic levels formed below the
parabola at the boottom of the Fermi sea. Notice that the
effective potential seen by the fermions deviates from the
original parabolic one only in the center of the trap due the
presence of the potential well created by the bosonic density.
Thus, the effective potential seen by the bosons is much deeper
and wider, due to the contribution of the wide fermionic density
and of their self-interaction. From this figure it is clear  that
two fermionic states have lowered their energies by adjusting
themselves  into the potential well created by the BEC giving rise
in this way to a {\it matter wave quantum dot} i.e. a localized
BEC with one or more fermionic levels trapped inside.
\begin{figure}[htb]
\centerline{
\includegraphics[width=4.2cm,height=4.2cm,clip]{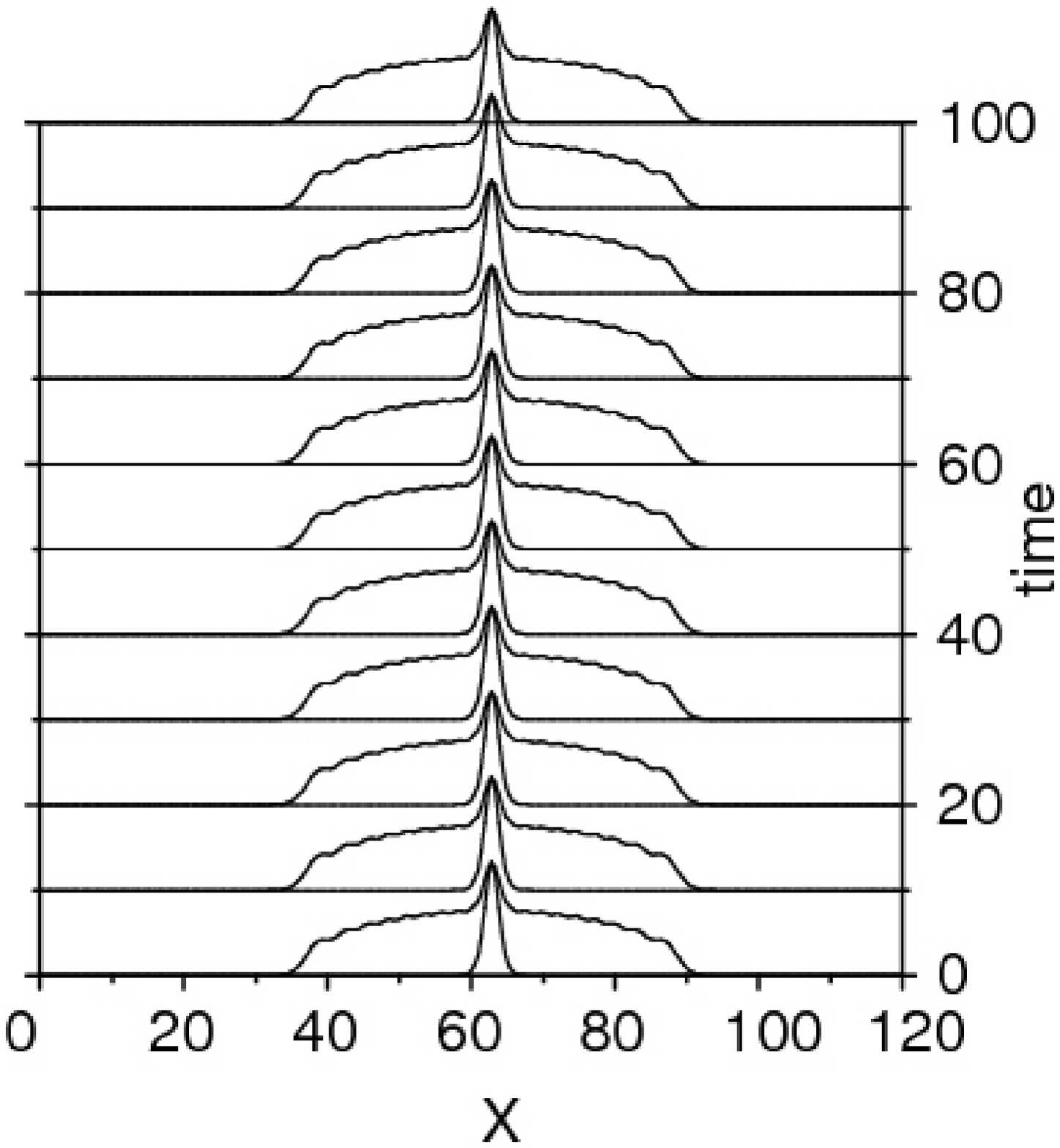}
\includegraphics[width=4.2cm,height=4.2cm,clip]{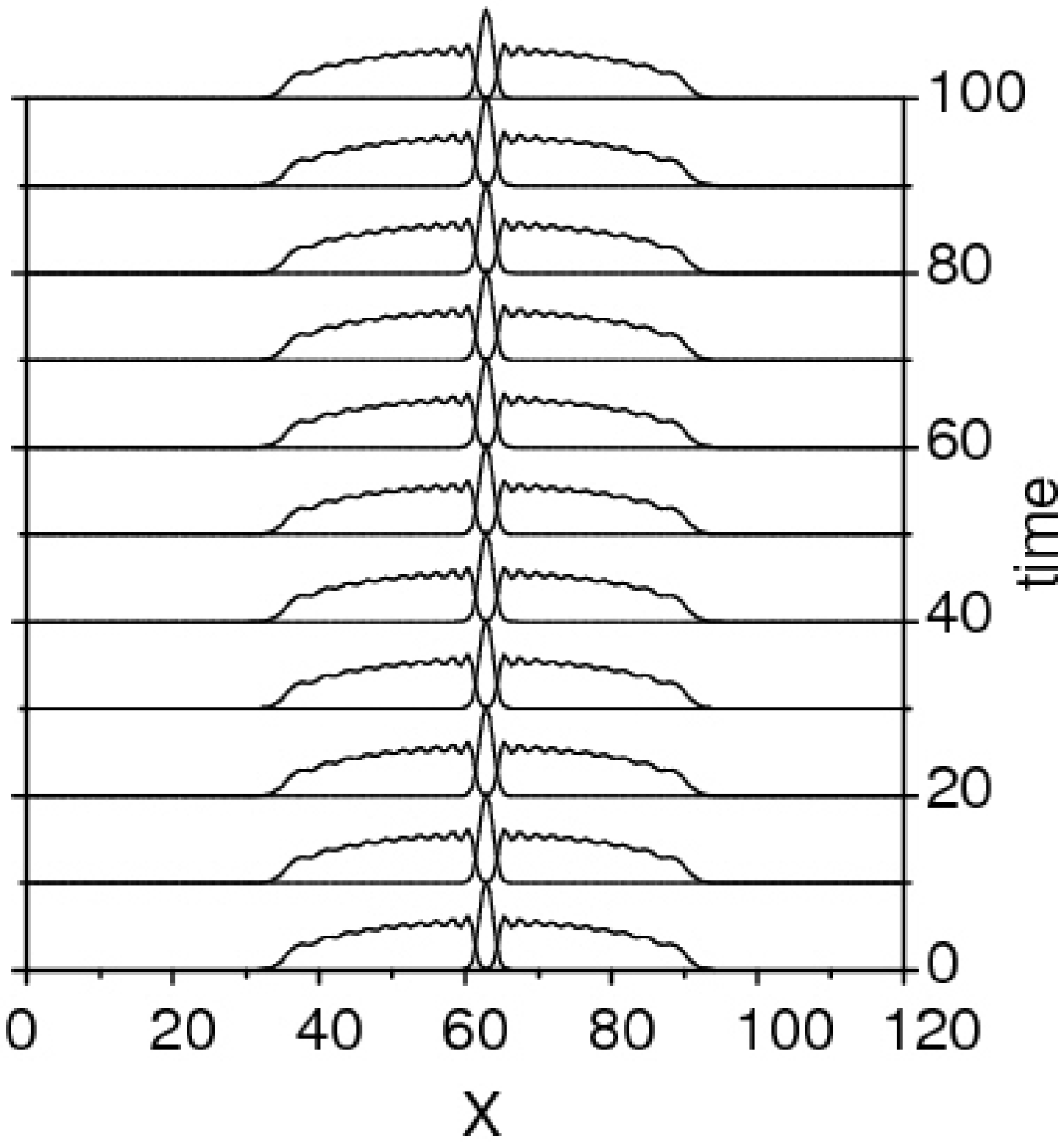}
} \caption{Time evolution of bosonic and fermionic densities in
Fig. \ref{fig1}a (left panel) and Fig. \ref{fig1}c (right panel)
obtained from the quasi-classical Eqs. (\ref{bose-fermi-eqns}) on
a line of length $L=40 \pi$. Space and time are in normalized
units. } \label{fig2}
\end{figure}
Notice that the two level system exists at the bottom of the Fermi
sea and are separated by a gap from other fermionic levels which
rapidly became equally spaced, as one would expect for a parabolic
trap. Also notice  that the ground state energy of the bosonic
part of the mixture is lowered inside the potential well
corresponding to the hump in the fermionic density and that
excited states, although lowering their energies,  remain almost
equally spaced (this is due to the fact that the effective
potential created by the fermions, except for the hump in the
middle, looks still parabolic).

It is remarkable that this situation remains stable under time
evolution, as one can see from the left panel of Fig. \ref{fig2}.
In Fig. \ref{fig1}c, we show the situation  for repulsive
Bose-Fermi interactions  ($\chi_{bf}>0$). From Fig. \ref{fig1}c we
see that in this case fermions are expelled from (instead of
attracted to) the region where bosons are spatially concentrated,
this increasing their kinetic energy and rising a pressure on the
condensate. Also notice  from Fig. \ref{fig1}d  that the
corresponding energy levels are shifted up in energy, much more
for bosons (which have an effective potential very deformed at the
bottom) than for the fermions (which see almost the original
parabolic trap but with a narrow hump in the middle). From this
figure it is clear that  the case of repulsive interactions
($\chi_{bf}>0$) is also of interest because, although not
producing trapping states, it induces correlations among the
fermions as is evident from the deviations from equally spaced
levels observed in the region across the Fermi surface in Fig.
\ref{fig1}d. This case can indeed be viewed as a {\it matter wave
anti-dot} and could be used to investigate effects of strong
correlations among fermions induced by the BEC .
\begin{figure}[htb]
\centerline{
\includegraphics[width=4.2cm,height=4.2cm,clip]{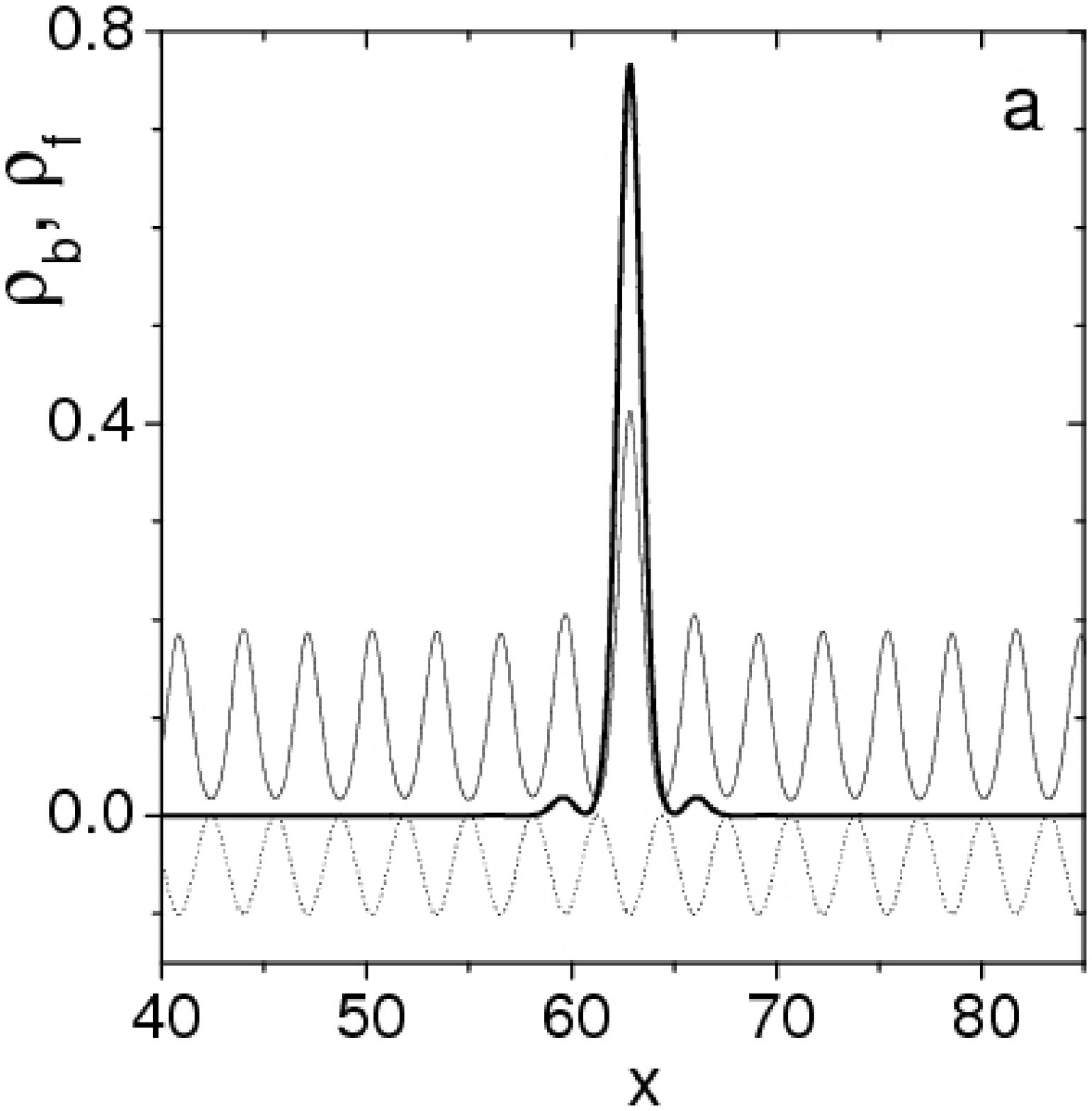}
\includegraphics[width=4.2cm,height=4.2cm,clip]{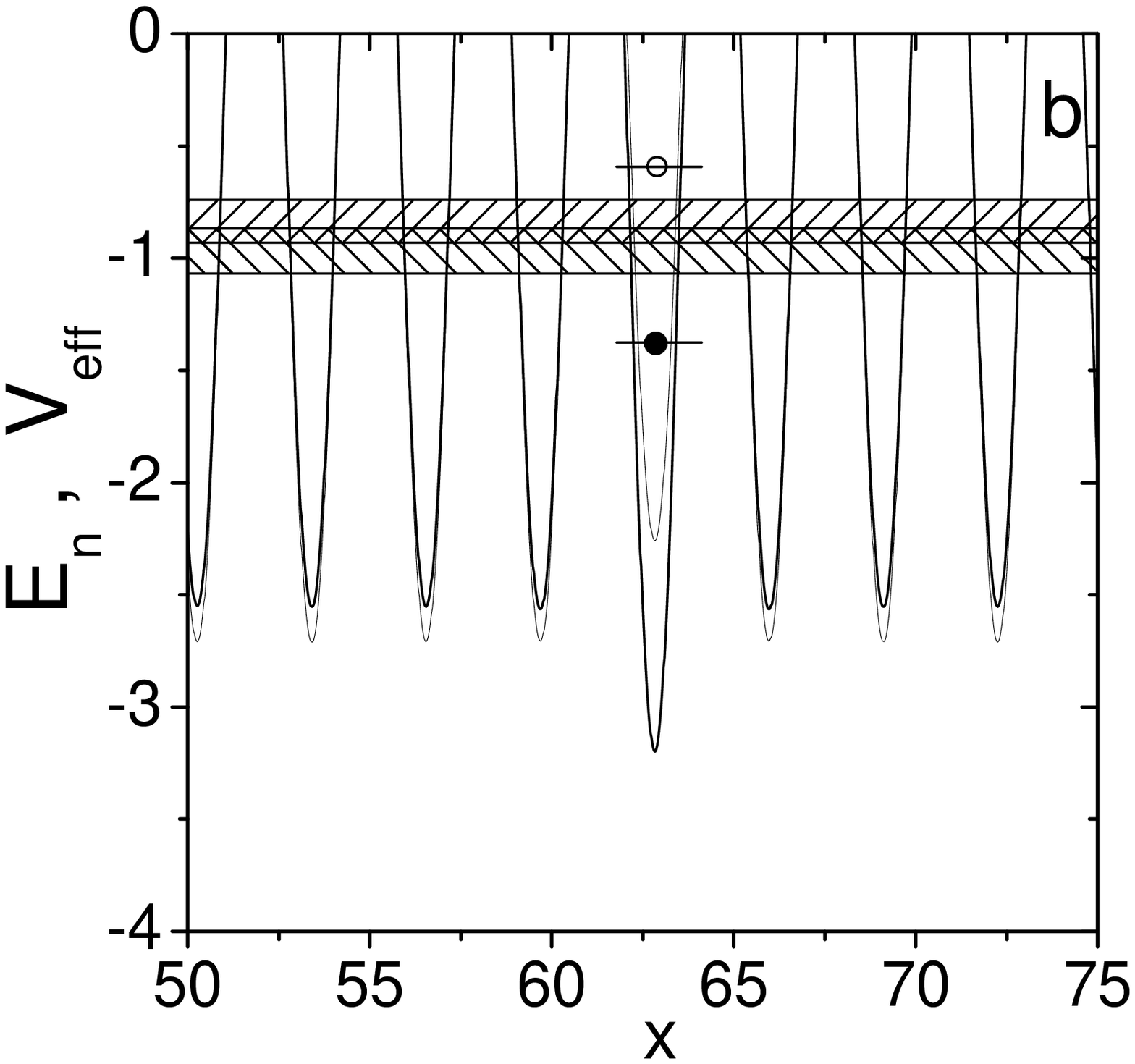}}
\centerline{
\includegraphics[width=4.2cm,height=4.2cm,clip]{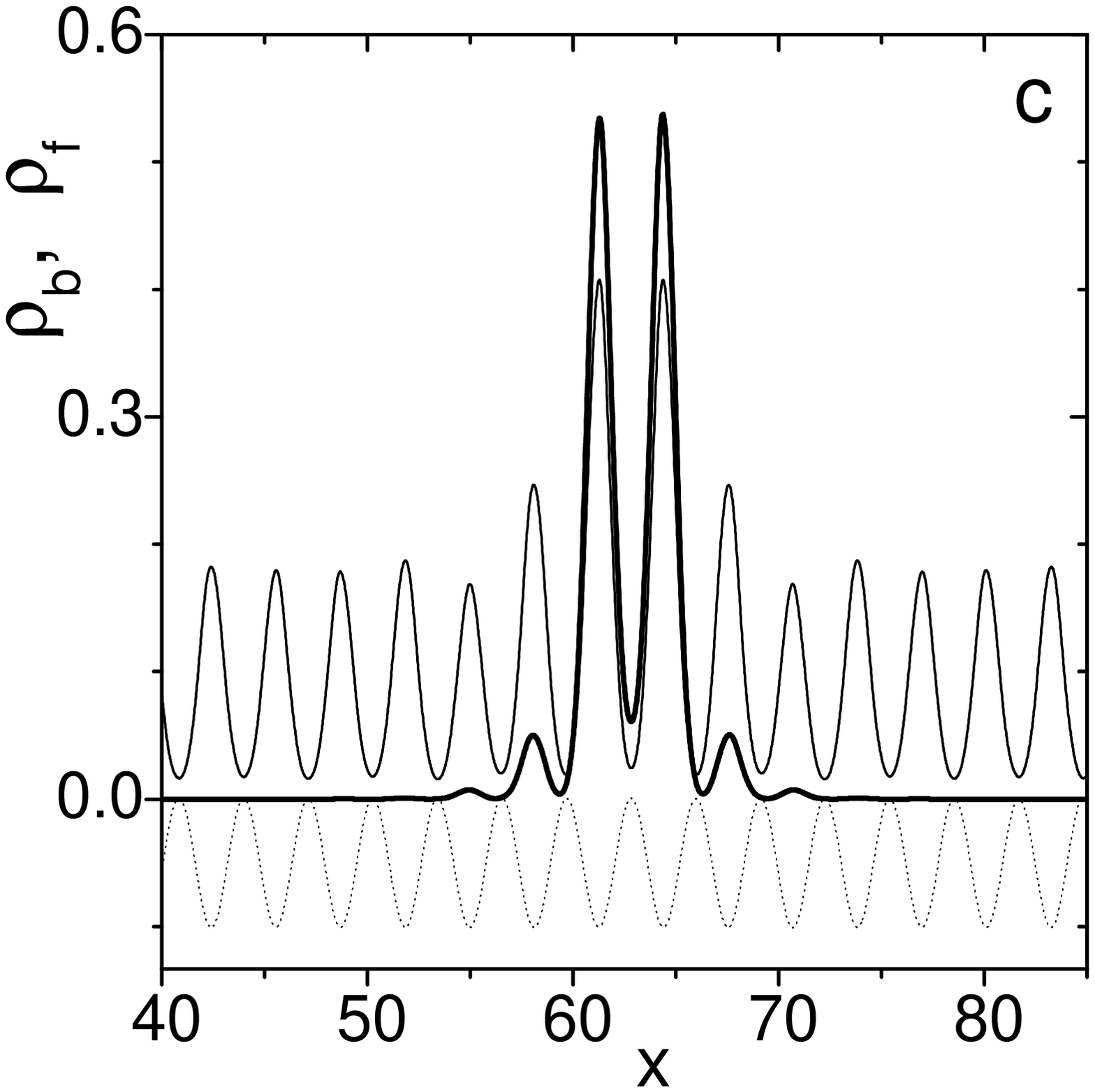}
\includegraphics[width=4.2cm,height=4.2cm,clip]{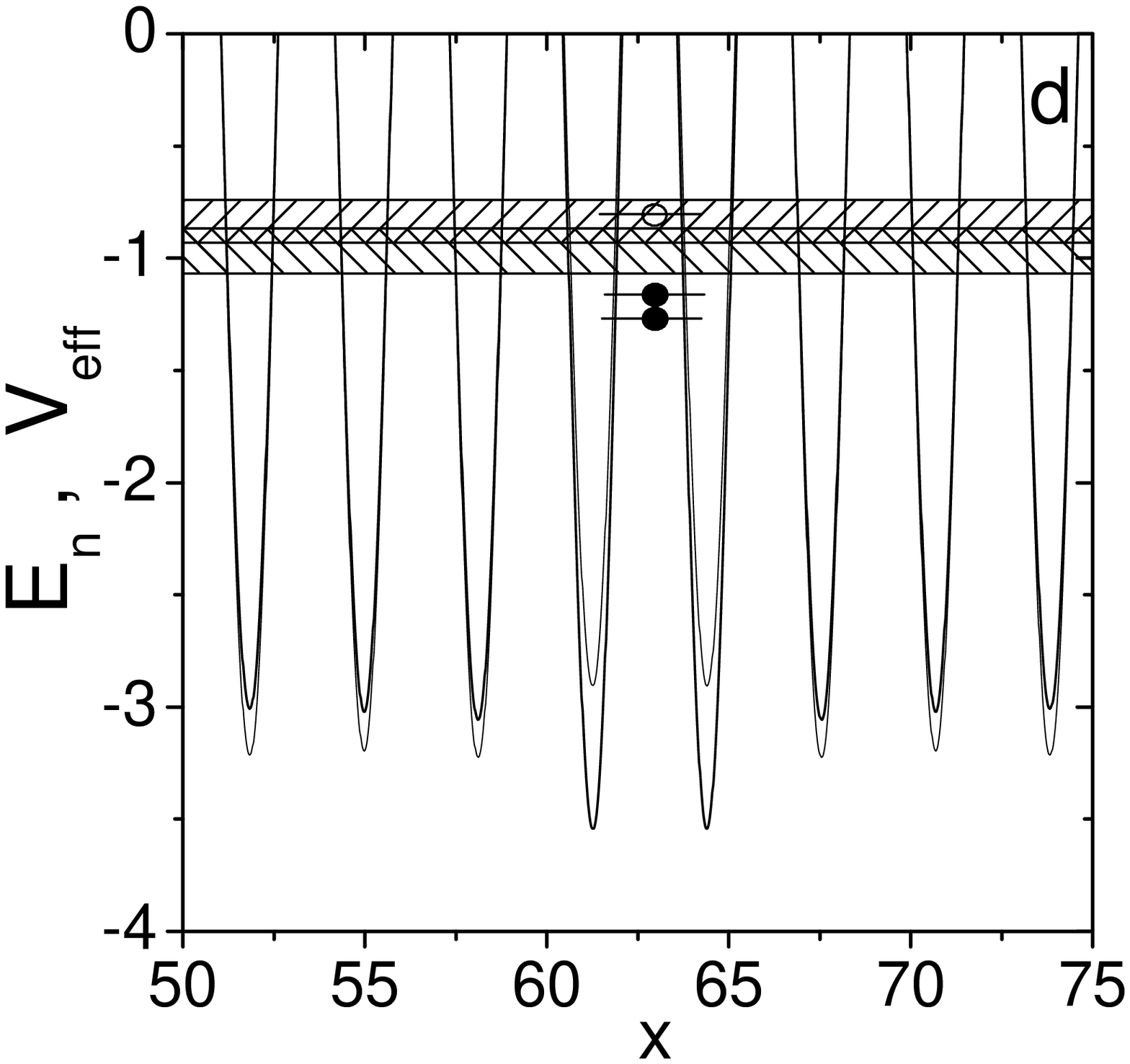}
}\caption{Panel (a). On-site symmetric densities for the case of
$\chi_{b}=1, \chi_{bf}=-1$, $N_f=20$  and normalized number of
bosons $N_b=1$ for an optical lattice of strength $\varepsilon=3$
(thick line refers to bosons). To identify the symmetry of the
state the OL is reported with an amplitude scaled by a factor of
$60$. Panel (b). Effective potentials (only lower part) and energy
levels of the Bose (thin line and open circles) and Fermi (thick
line and filled circles) mixture corresponding to the densities in
panel(a). Lowest bosonic band (shadow region with negative slope)
and lowest fermionic band (shadow region with positive slope) have
also been shown. The fermionic band is half filled. Panels (c) and
(d): the same as panels (a),(b), but for the inter-site symmetric
case with $N_b=1.47$ . Plotted quantities are in normalized
units.} \label{fig3}
\end{figure}
\section{Bose-Fermi mixtures in optical lattices}

In this section we consider a trap potential of the form
$V_{trap}=V_0 \cos (2 k_L x) $, as a  model for an optical
lattice. We assume $m_b=m_f=m$, $V_{ext}^b= V_{ext}^f=V_{trap}$
and normalize Eqs. (\ref{bose-fermi-eqns}) by measuring space in
units of $k_L^{-1}$, time in units of $\hbar/E_r$ ($E_r=\hbar^2
k_L^2/2m$ is the lattice recoil energy) and rescaling
wavefunctions according to $\psi_b \rightarrow (g_{{b}_0} N_b /2
\pi E_r {a_{\perp}}^2)^{1/2} \psi_b$, $\psi_f\rightarrow
(g_{{bf}_0}/2\pi E_r {a_{\perp}^2})^{1/2})^{1/2}) \psi_f $. This
leads to the same normalized equations  (\ref{norm-eqns}) but with
$V=\varepsilon \cos(2x)$, where $\varepsilon=V_0/E_r$. With this
normalization the relation between the dimensionless number of
bosons $N_b$ and the physical one is $N_p=(k_L
a_{\perp}^2/a_{b})N_b$ (typical values are in the range of
$10^3-10^5$ atoms, for $a_{\perp}\approx 10^{-6} m, k_L=10^7
m^{-1}, a_{b}= 1 \div 50 \cdot 10^{-10}m$). In the following we
show the existence of gap-solitons in BFM and suggest them as
matter wave quantum dots (anti-dots) (the anti-dot case  will be
discussed elsewhere \cite{msalerno}).
\begin{figure}[htb]
\centerline{
\includegraphics[width=4.3cm,height=4.8cm,clip]{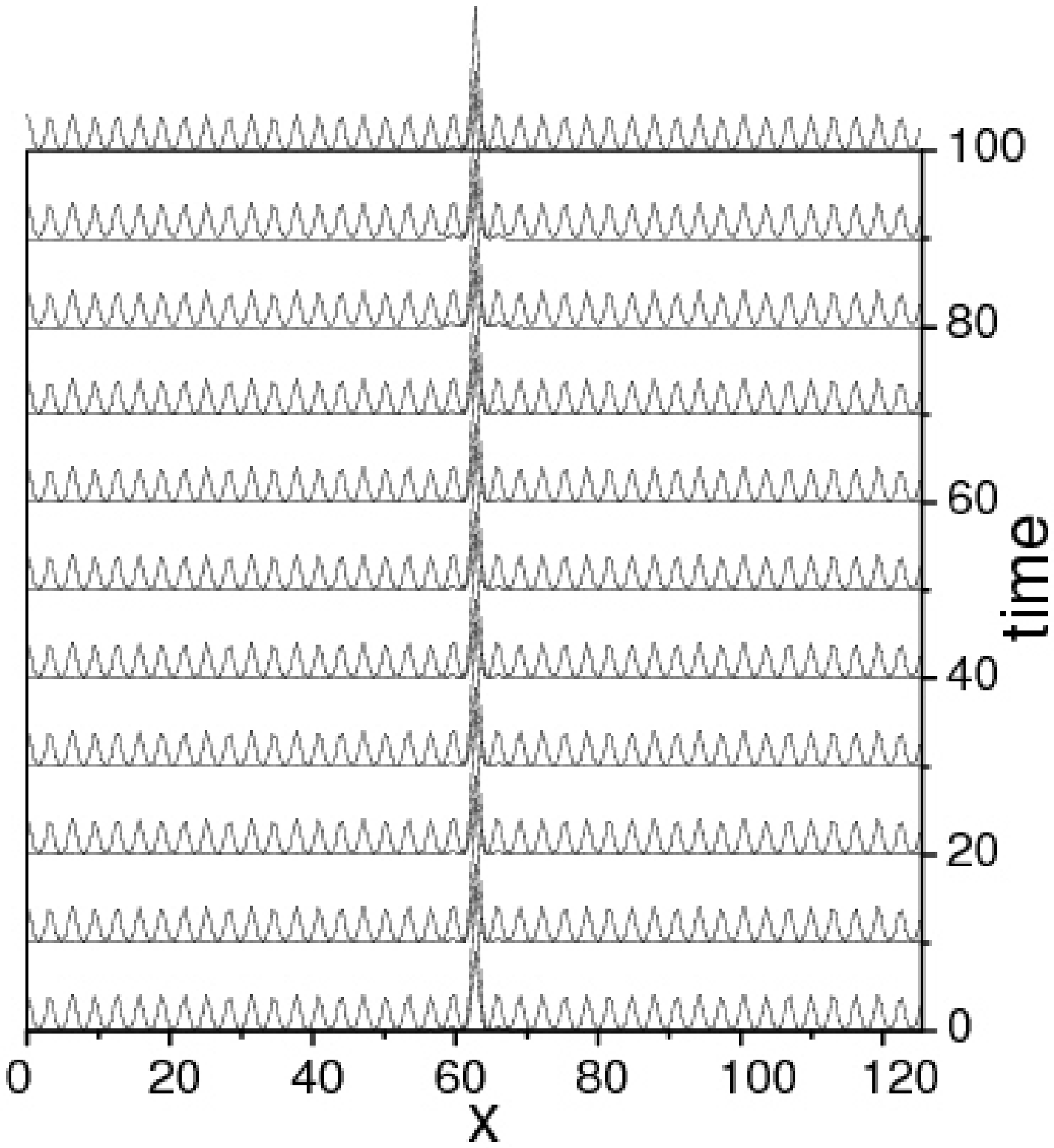}
\includegraphics[width=4.3cm,height=4.8cm,clip]{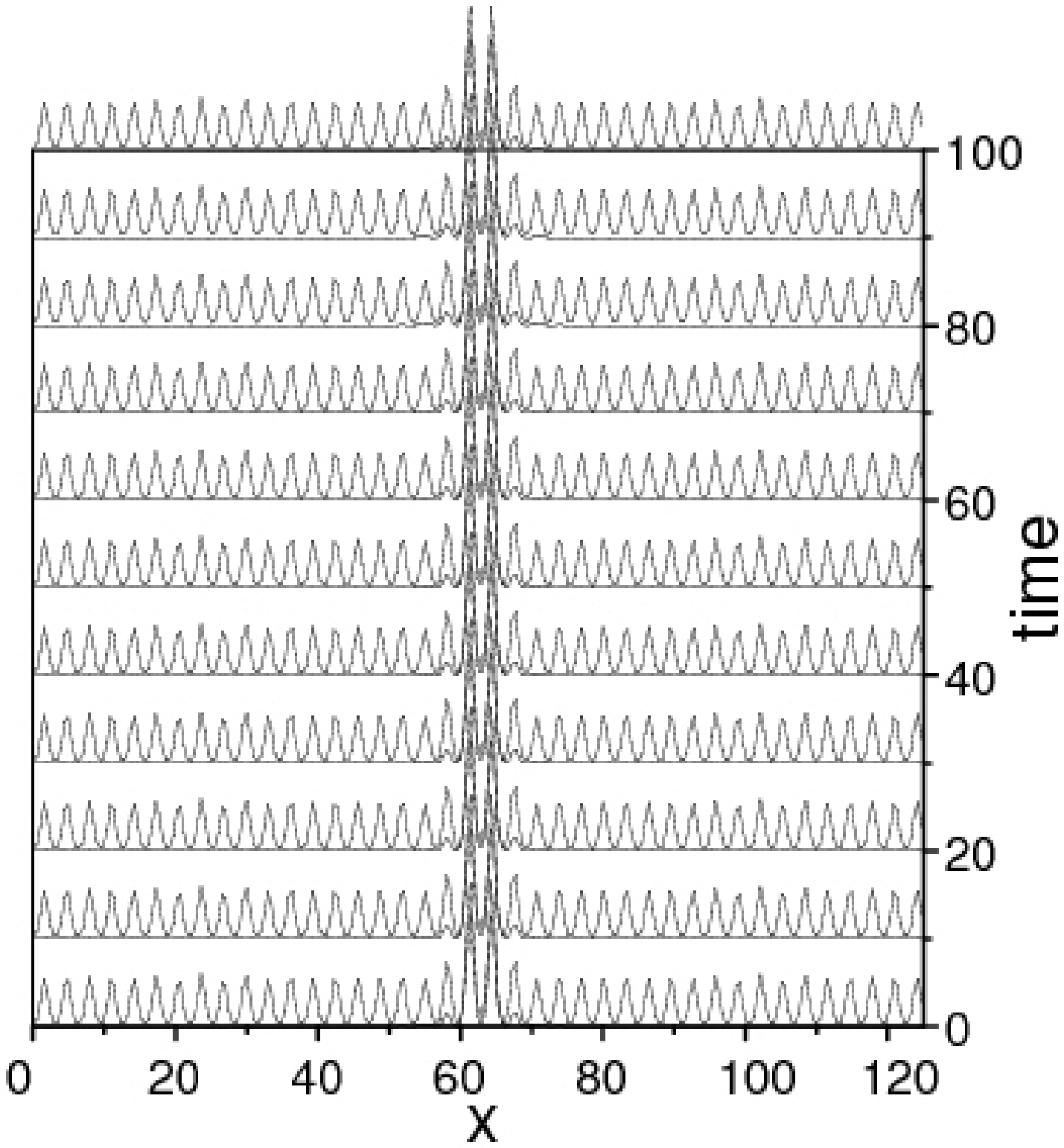}}
\caption{Time evolution of the OS (left panel) and IS (right
panel) BFM densities in Fig. \ref{fig3}a and Fig. \ref{fig3}b
obtained from direct numerical integrations of  Eqs.
(\ref{bose-fermi-eqns}) on a line of length $L=40 \pi$. Space and
time are in normalized units.} \label{fig4}
\end{figure}

To this regard we consider BFM with repulsive bosonic-bosonic
$\chi_{b}>0$ and attractive bosonic-fermionic $\chi_{bf}<0$
interactions. In Fig. \ref{fig3}a the bosonic and fermionic
densities corresponding to an onsite symmetric (OS) gap soliton,
are depicted (notice that the soliton is symmetric around a
minimum of the potential). We see that while the bosonic density
is localized in almost a single potential well, the fermionic
density is very extended in space and presents a hump in
correspondence of the BEC. In Fig. \ref{fig3}b the effective
potentials, the sketch of the band structure and the energy levels
of localized states for both bosons and fermions are shown. Notice
that the energy of the OS gap soliton lies in the gap between the
first two bands as expected for repulsive bosonic interactions,
while the fermionic level responsible for the density hump in Fig.
\ref{fig3}a, lies below the lowest energy band at the bottom of
the Fermi sea. As for the parabolic case, gap soliton states with
trapped fermionic levels inside can be seen a matter wave
realization of quantum dots.

More complicated gap soliton states (or arrays of gap solitons)
can also be formed. This is shown in Fig. \ref{fig3}c for the case
of a two hump BEC condensate obtained as inter-site symmetric (IS)
gap-soliton (notice that the state is symmetric around a maxima
instead than a minima of the periodic potential). We see that also
in this case the fermionic density increases in correspondence
with the two bosonic humps and two fermionic levels are formed in
the corresponding effective potential (see Fig. \ref{fig3}d). We
have checked that both the OS and the IS modes are very stable
under time evolution (see Fig. \ref {fig4}).

\begin{figure}[htb]
\centerline{
\includegraphics[width=4.2cm,height=4.2cm,clip]{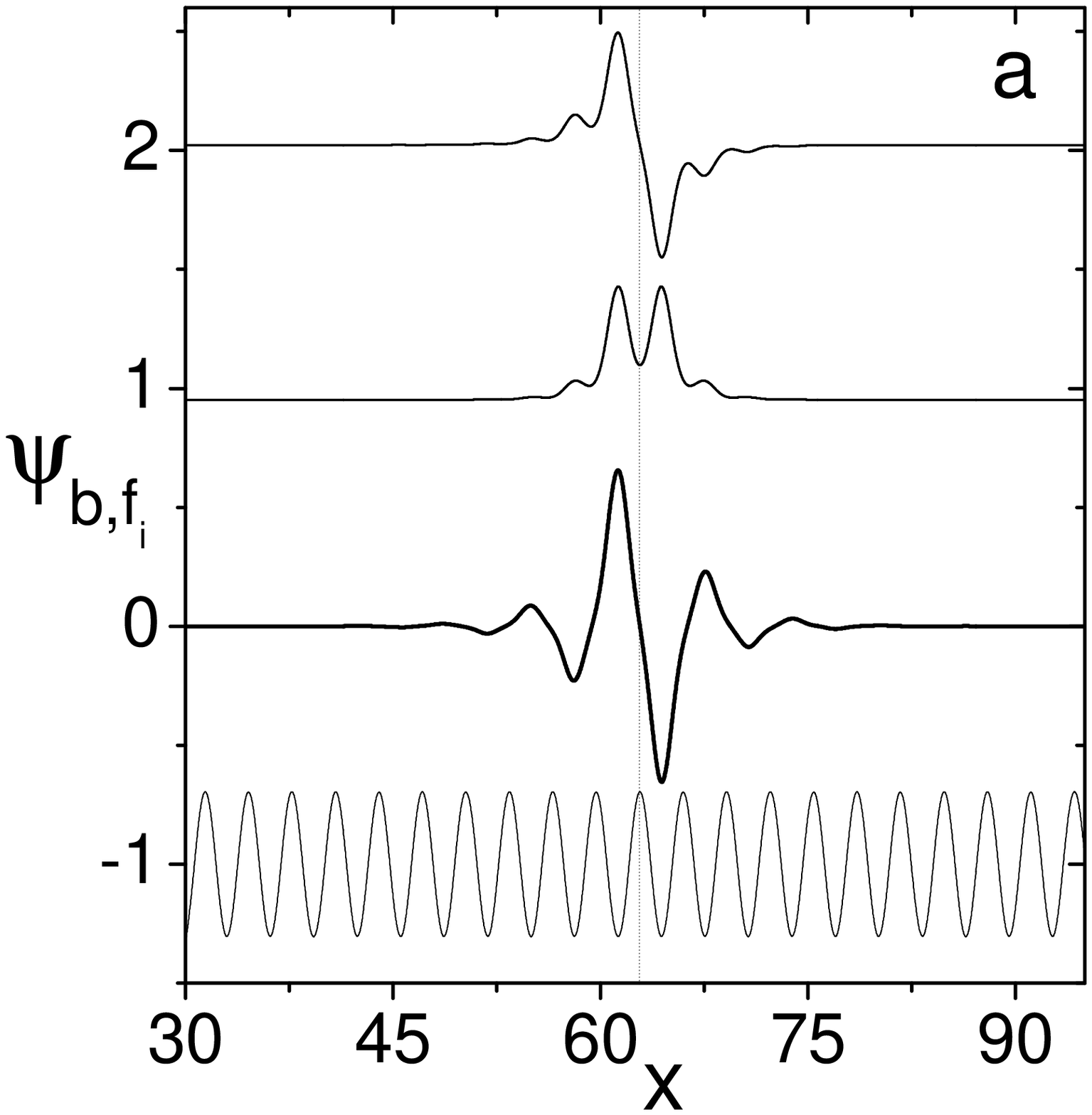}
\includegraphics[width=4.2cm,height=4.6cm,clip]{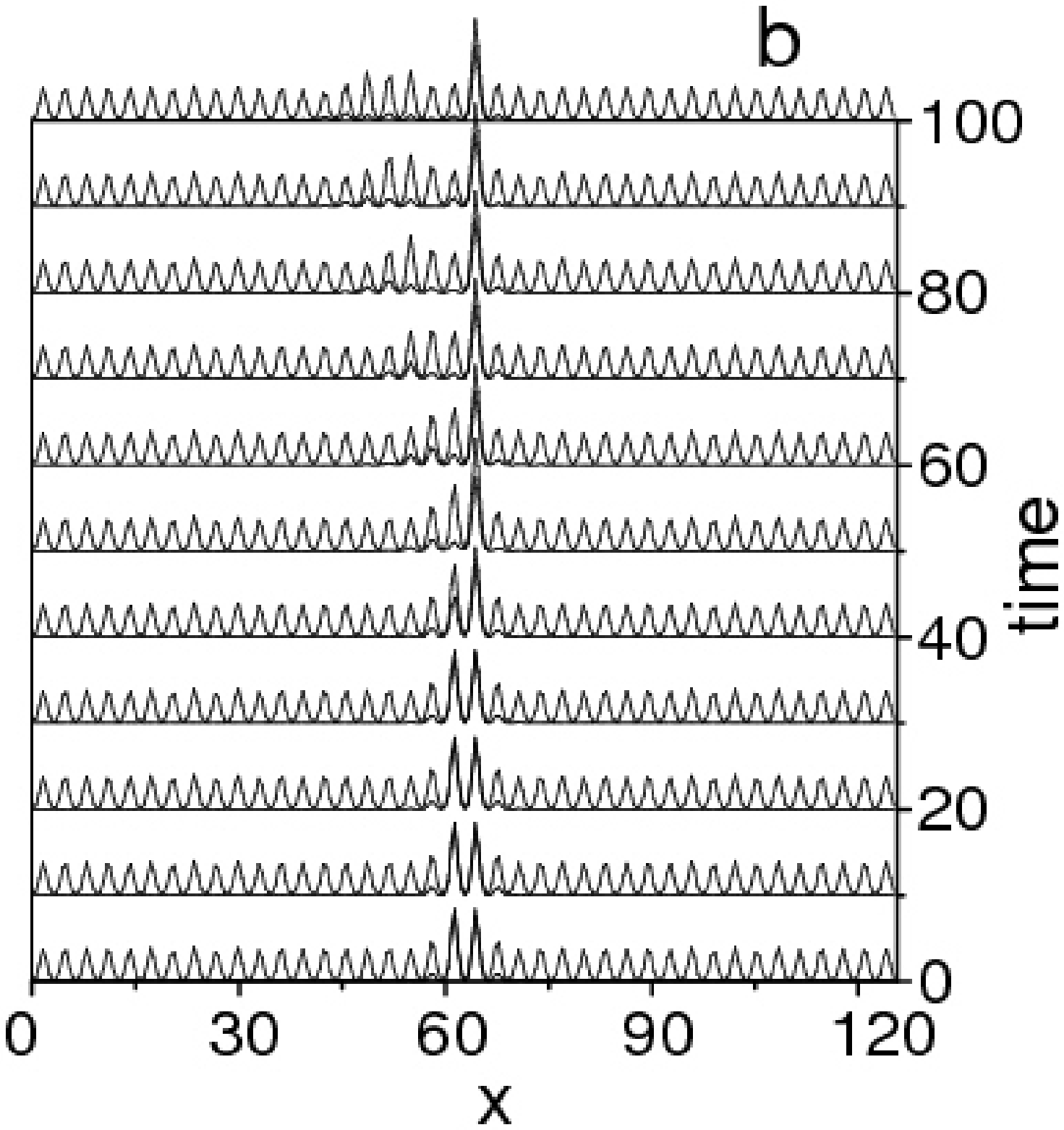}}
\centerline{
\includegraphics[width=4.2cm,height=4.2cm,clip]{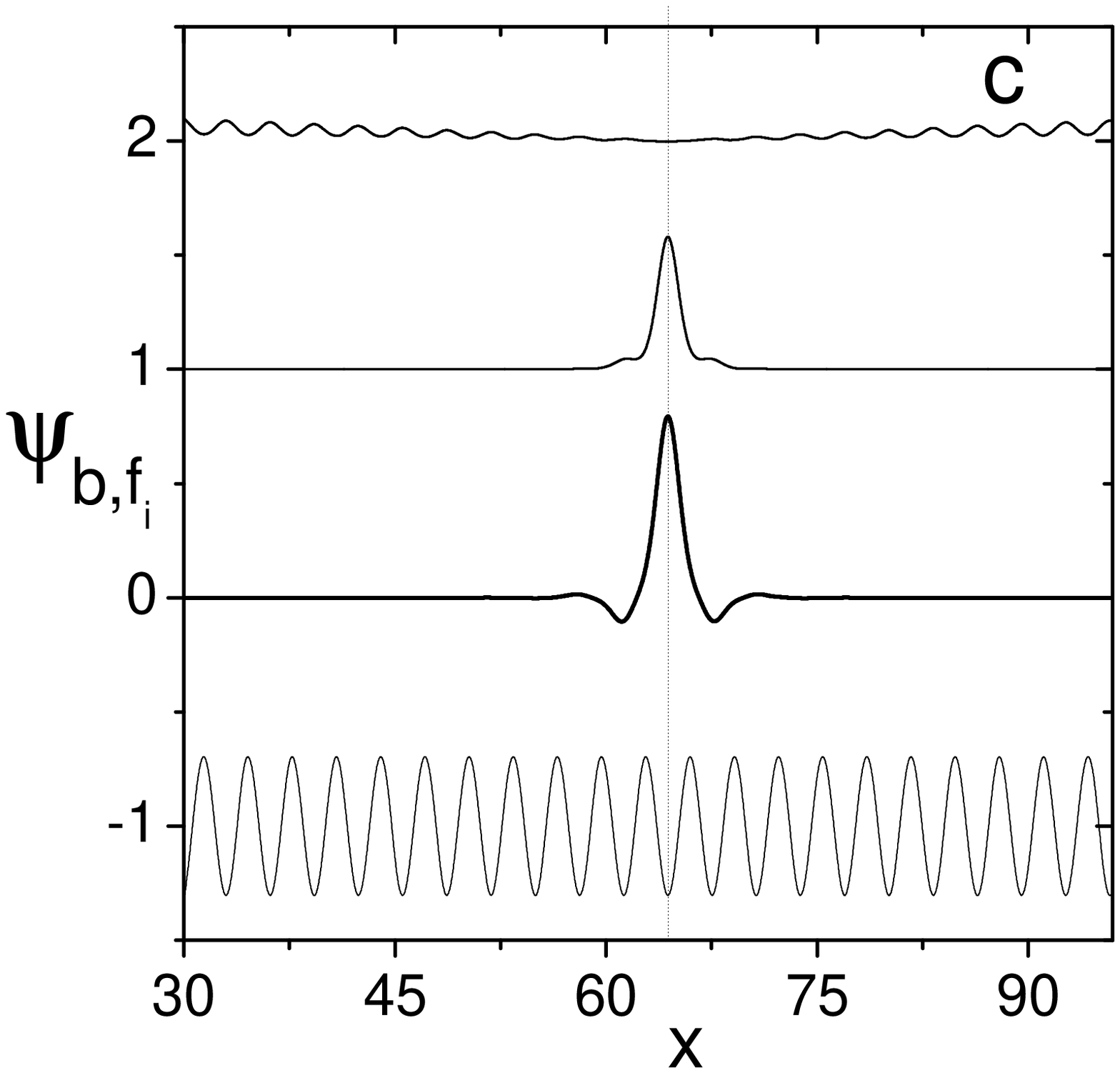}
\includegraphics[width=4.2cm,height=4.6cm,clip]{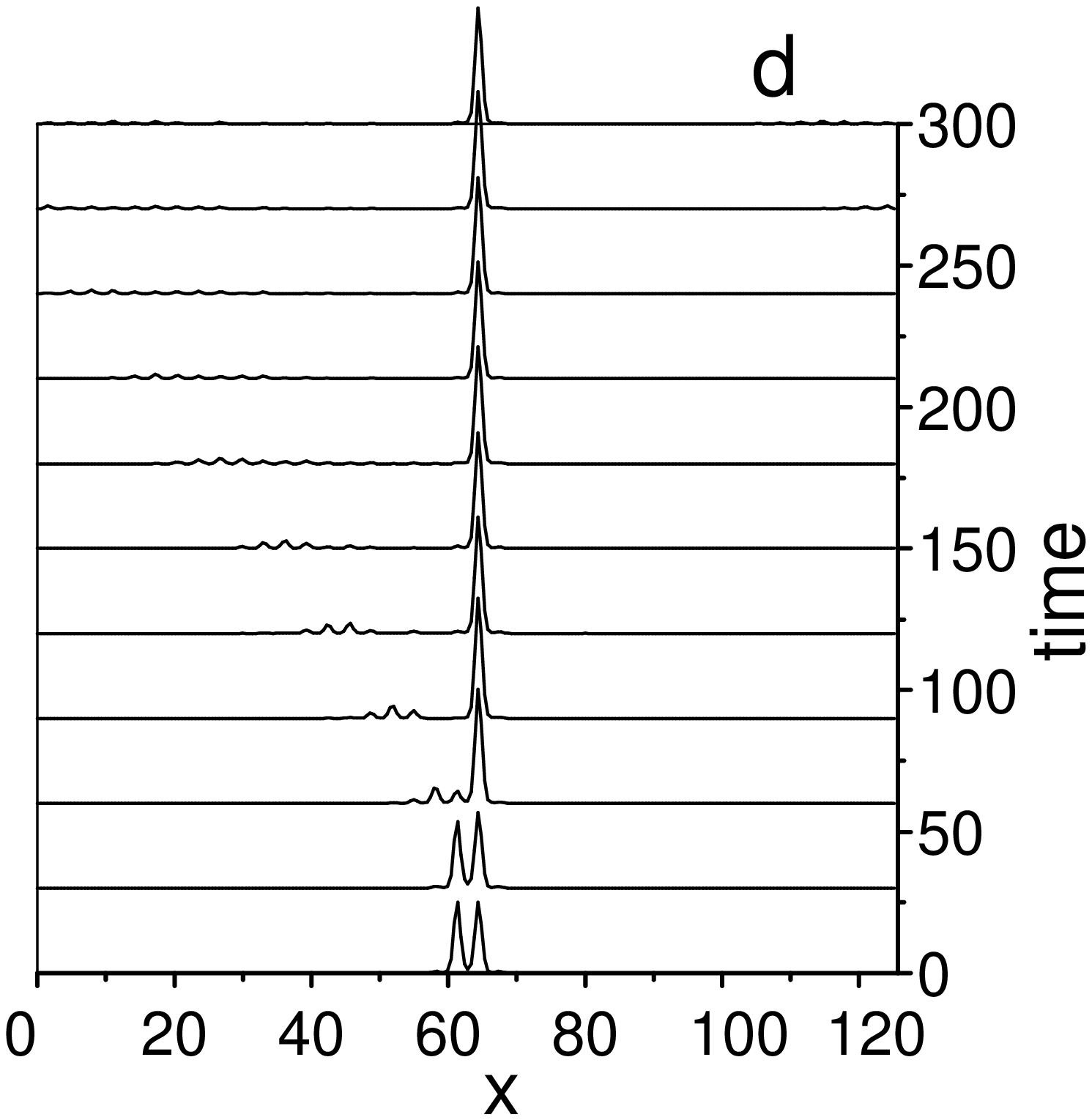}
} \caption{{\bf (a)} Wavefunction of the intersite-asymmetric (IA)
gap-soliton (thick line) plotted together with the wavefunctions
of the lowest  and first excited fermionic states trapped in the
condensate (second and first curve from the top, respectively).
The periodic potential, scaled by a factor 10 for graphical
convenience, is also shown with a thin line at the bottom.
Parameter values are the same as in Fig. \ref{fig3} with $N_b=1$.
{\bf (b)} Time evolution of the bosonic and fermionic densities of
the metastable IA gap-soliton obtained from the quasi-classical
Eqs. (\ref{bose-fermi-eqns}) on a line of length $L=40 \pi$.
Panels {\bf (c)}. Same as panel (a) but for the OS state obtained
after the transition. Panel {\bf (d)}. Time evolution of the
wavefunction of the lowest energy state inside the condensate.
Plotted quantities are in normalized units.} \label{fig5}
\end{figure}

Gap- solitons with different symmetries with respect to the OL,
such onsite-asymmetric (OA) and intersite-asymmetric (IA), can
also exist  in BFM  but they appear to be metastable under time
evolution. In  Fig. \ref{fig5}a we show a gap-soliton of type IA
together with two fermionic states trapped in the condensate. In
panel (b) of this figure we depict the time evolution of the
bosonic and fermionic densities of the IA gap-soliton as obtained
from direct numerical integration of Eqs. (\ref{bose-fermi-eqns}).
We see that after a time $t\approx 60$ the IA state decays into
the stable OS wavefunction depicted in Fig. \ref{fig5}c. Notice
that the IA-OS transition also induces changes in the fermionic
states. In particular we see that the first excited  state
asymptotically decays into the Bloch state at the bottom of the
fermionic band (see top curve of Fig.\ref{fig5}c), while the
lowest energy fermionic state remains well localized inside the
gap soliton (see the time evolution reported in panel (c)). The
IA-OS transition in Fig. \ref{fig5}c occurs therefore with the
expulsion of one fermion from the condensate and with a change of
symmetry of the wavefunction of the other fermion from IA to OS
type (see second curves from the top in panels (a) and (c)).
Similar phenomena occur for other metastable gap-solitons states
existing in higher energy gaps and for different sign combinations
of the interatomic interactions.

Also notice from Fig. \ref{fig5}b that the IA-OS transition occurs
with emission of matter from the condensate meaning that the final
OS gap soliton state contains less matter than the initial state.
This implies that the corresponding effective potential for the
fermions is weaker after the transition  and therefore less
effective to hold a second bound state in the condensate. As the
boson-fermion (attractive) interaction and the number of bosons in
the gap soliton are increased, we find that more fermionic levels
are able to enter the condensate.

An interesting problem to investigate is how a matter wave q-dot
(gap soliton) gets progressively filled with fermions as the
product $\chi_{bf}N_b$ is increased. To this regard we consider OS
bosonic gap solitons with attractive boson-fermion interaction and
with both repulsive and attractive boson-boson interactions.
\begin{figure}[htb]
\centerline{\hskip-.5cm
\includegraphics[width=4.4cm,height=4.8cm,clip]{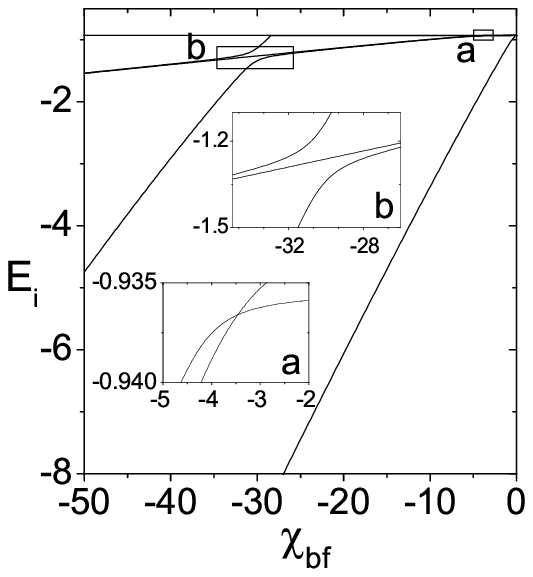}
\includegraphics[width=4.4cm,height=4.8cm,clip]{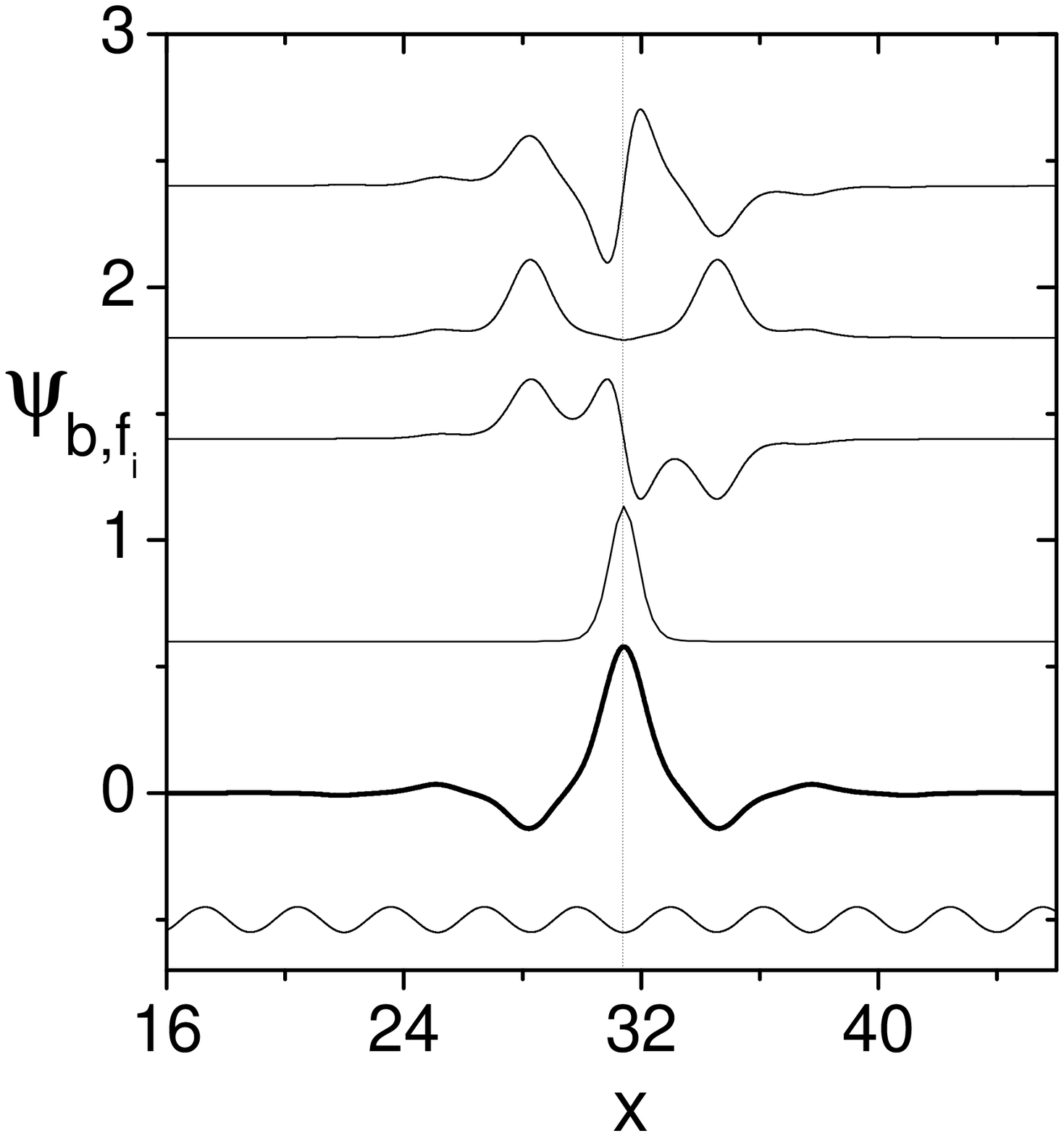}
} \caption{Left panel. Energy levels of the fermionic bound states
inside the condensate as a function of $\chi_{bf}$, for parameter
values $\chi_{b}=1, \varepsilon=3, N_b=1,  N_f=20$. The top
horizontal line represents the bottom of the fermionic band while
the insets are enlargement of rectangles a,b, where level crossing
and avoid crossing occur, respectively. Right Panel. Wavefunctions
of the gap soliton (thick line) and of trapped fermionic states
(ordered by decreasing energy from the top) at $\chi_{bf}=-30$.
Other parameters are fixed as in the right panel. The vertical
dotted line and the periodic potential at the bottom (scaled by a
factor $60$) were drawn to identify the symmetry of the
wavefunctions. A vertical offset among curves was introduced to
avoid overlapping. All plotted quantities are in normalized
units.} \label{fig6}
\end{figure}
In the left panel of Fig. \ref{fig6} we show the energy of the
fermionic bound states inside the condensate as a function of
$\chi_{bf}$ for the case of a repulsive BEC ($\chi_b>0$) with a
fixed number of atoms. We see that the presence of an arbitrary
small attraction between bosons and fermions is enough to trap at
least one fermion in the gap soliton (notice that the energy of
the  fermion is below the edge of the lowest fermionic band, shown
as a horizontal line in the figure). This is a consequence  of the
one dimensional nature of the effective potential (for BEC in
higher dimensions a threshold in $\chi_{bf}$ is expected  for the
entering of the first fermion). From the inset (a) of the figure
we see that two fermionic levels of almost equal energy enter the
gap soliton at $\chi_{bf}\approx -3.5$. Notice that these levels
come from extended (Bloch) states inside the fermionic  band and
are pulled below the lower band edge by the attractive
boson-fermion interaction. The inset also shows that the two
levels undergo a level crossing after which they proceed together
as quasi degenerated states until a collision with a fourth level
occurs. Notice that the last level detaches from the bottom of the
fermionic band at $\chi_{bf}\approx -28.6$ and, in contrast with
the previous case, it undergoes an avoid level crossing with the
quasi degenerated states at $\chi_{bf}\approx -30$ (see inset
(b)). Also notice that the avoid crossing occurs with an exchange
of degeneracy, i.e. the newly entered level, originally non
degenerate, becomes quasi degenerate after the interaction while
the lowest energy level of the quasi degenerate pair becomes non
degenerate after the interaction.
\begin{figure}[htb]
\centerline{
\includegraphics[width=4.3cm,height=4.8cm,clip]{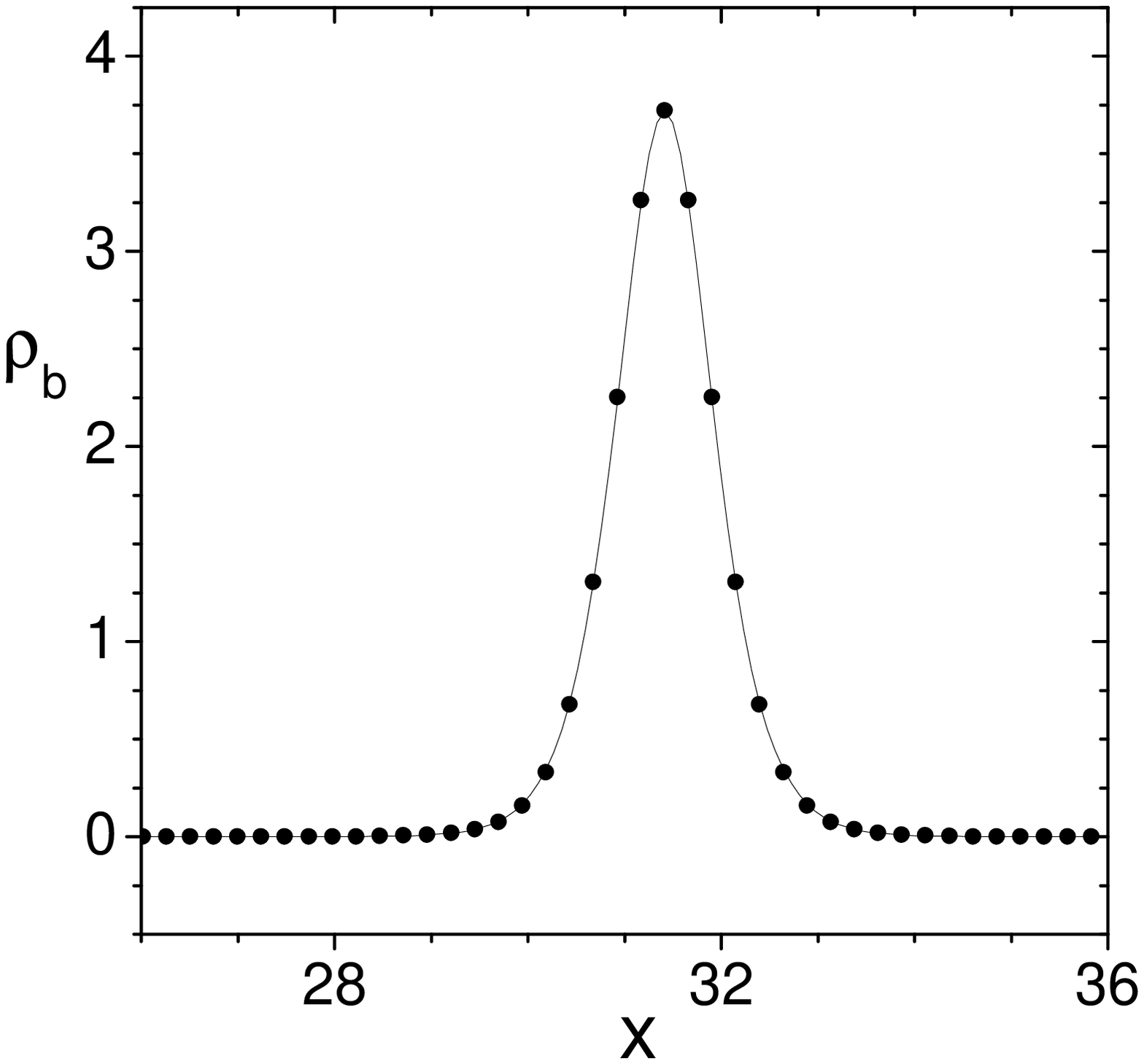}
\includegraphics[width=4.3cm,height=4.8cm,clip]{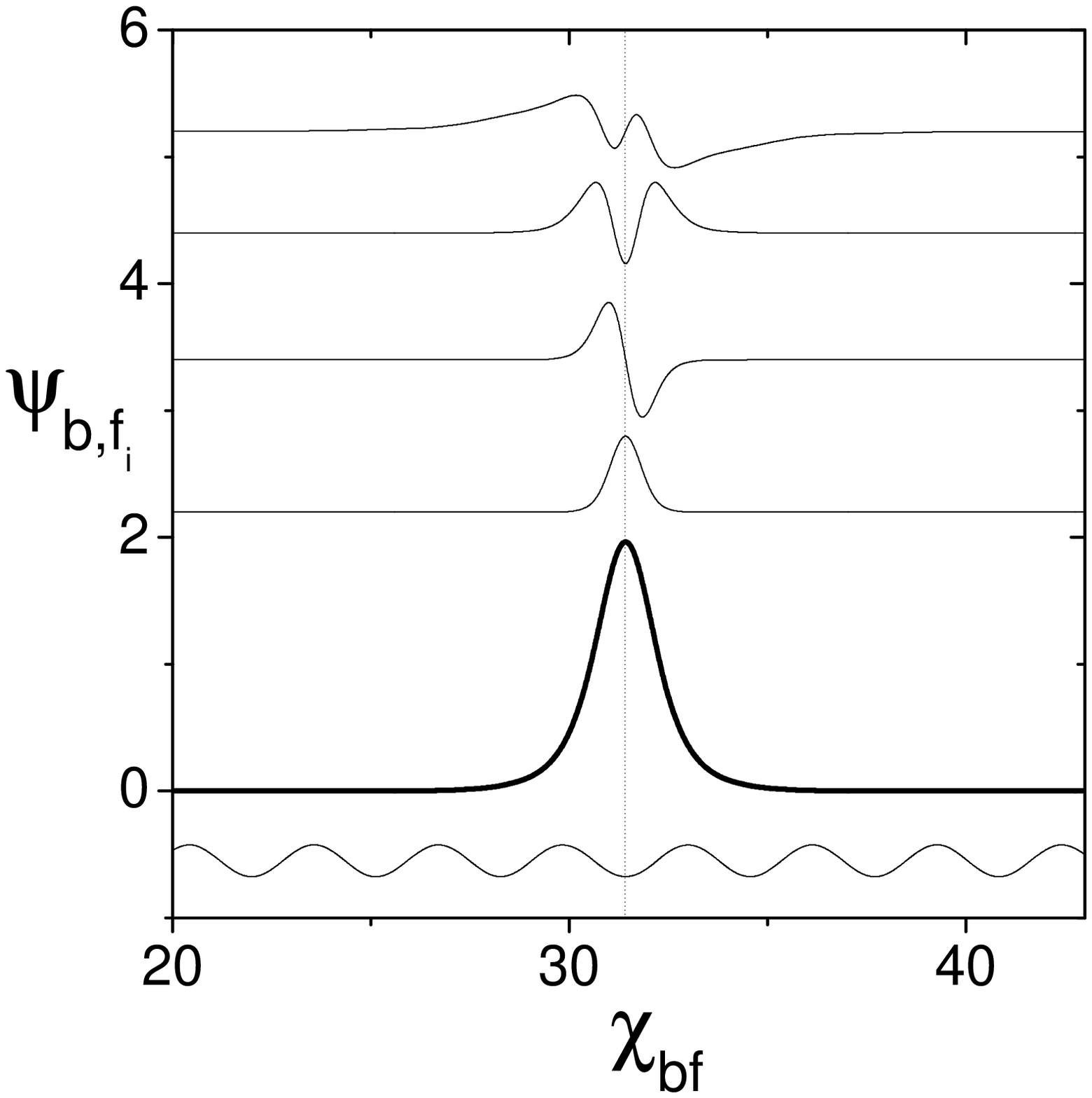}}
\caption{Bosonic  density for attractive boson-boson interaction
in an OL of strength $\varepsilon=0.5$. The continuous curve
refers to the potential in Eq. \ref{pt} with $a=0.66$ while the
dots represent self-consistent numerical results. Other parameters
are fixed as $\chi_b=-1, \chi_{bf}=-10, N_b=4.9, N_f=20$. Right
panel. Gap soliton (thick line) and trapped fermionic
wavefunctions ordered by decreasing energy from the top.
Parameters are fixed as in the left panel. The vertical dotted
line and the periodic potential at the bottom (scaled by a factor
of $4$) are drawn to identify the symmetry of the wavefunctions. A
vertical offset among curves was introduced to avoid overlapping.
All plotted quantities are in normalized units.} \label{fig7}
\end{figure}

This behavior is in agreement with a general theorem of quantum
mechanics, holding for hamiltonians depending on a parameter (
with eigenvalues functions of the parameter) according to which
the level crossing occurs between levels of different symmetry and
avoid crossing between states with the same symmetry. In the
present case the wavefunctions of the fermionic levels inside the
BEC can be either symmetric or asymmetric with respect to the OL
site around which the  gap soliton is centered. Moreover, the
symmetry alternates between consecutive levels, the lowest state
being always of OS type. This is seen from the right panel of Fig.
\ref{fig6} where we have depicted the gap soliton found at
$\chi_{bf}=-30$ with the four fermionic states trapped inside.
Notice  that while the second and third bound states have
different symmetries, and therefore their levels can cross, the
second and fourth excited wavefunctions have the same symmetry and
therefore their levels must give rise to an avoid crossing when
approaching each other. This general rule for the crossing of the
fermionic levels inside a gap soliton is in full agreement with
the self-consistent numerical calculations shown in insets (a)-(b)
of Fig. \ref{fig6}.

Analytical estimates for the energy levels and for the number of
fermions in the condensate can be obtained for the case of an
attractive BEC in shallow OL. To this regard we remark that for
repulsive boson-boson interactions the bosonic density cannot be
localized into a single potential well (due to the tunneling of
matter into adjacent wells induced by the repulsive interaction),
so that satellites around a main peak always appear (see the right
panel of Fig. \ref{fig6}). In this case the corresponding
effective potential in the fermionic Schr\"odinger equation gives
rise to a spectral problem which is analytically difficult to
solve. For attractive boson-boson interactions, however, the
bosonic density may have no satellites peaks (due to the
attraction)  and can be easily approximated by integrable
potentials (this is especially true when the OL is shallow).

In the following we consider the case of an attractive BEC with
$\chi_b=-1$ in  an OL of strength $\varepsilon=0.5$ and
approximate the bosonic density with a P\"oschl-Teller potential
(soliton potential) of the form
\begin{equation}
\rho_b \approx \frac{N_b}{2 a} \cosh^{-2} (\frac xa). \label{pt}
\end{equation}

\begin{figure}[htb]
\centerline{
\includegraphics[width=4.5cm,height=4.5cm,clip]{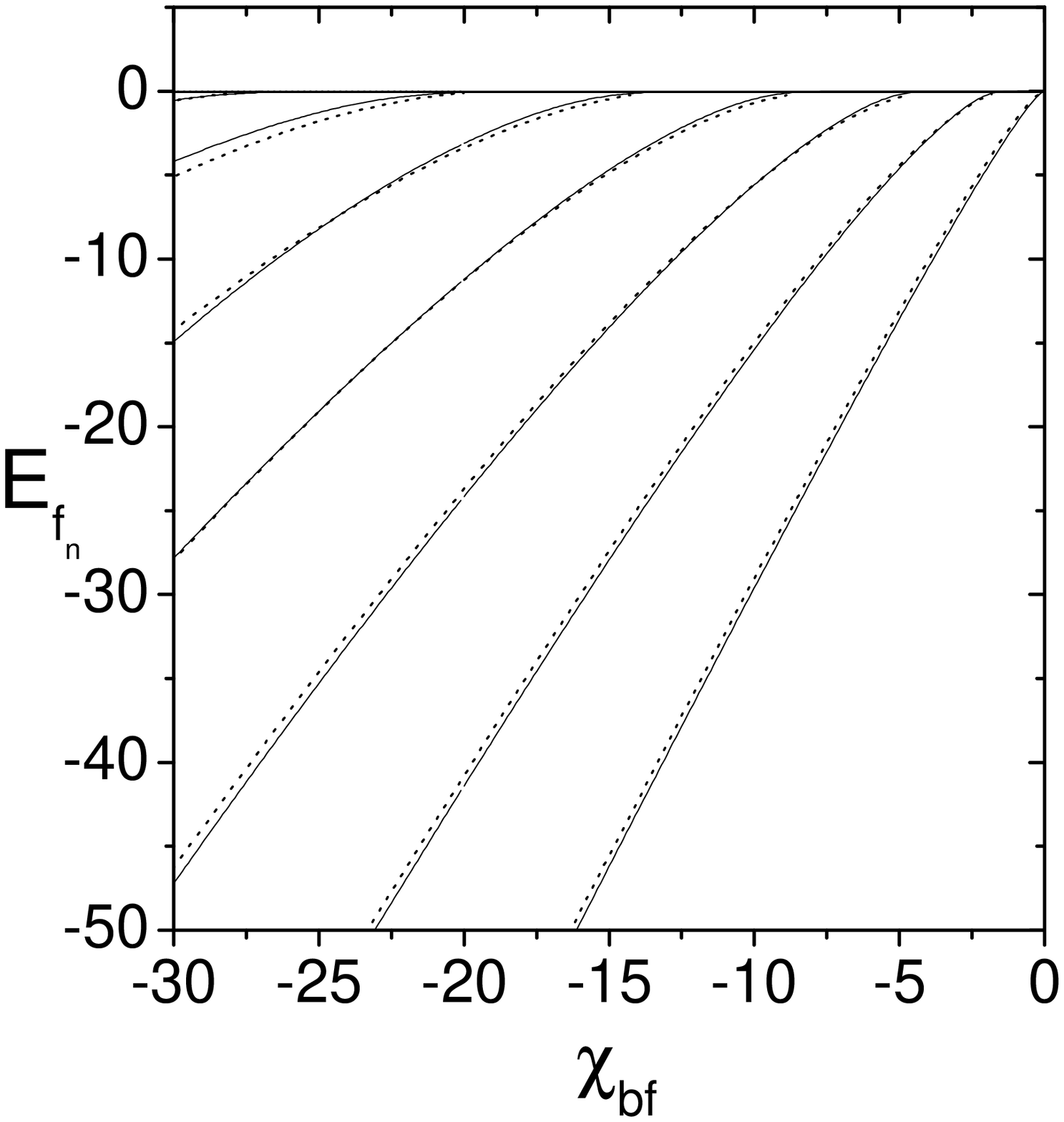}
\includegraphics[width=4.5cm,height=4.5cm,clip]{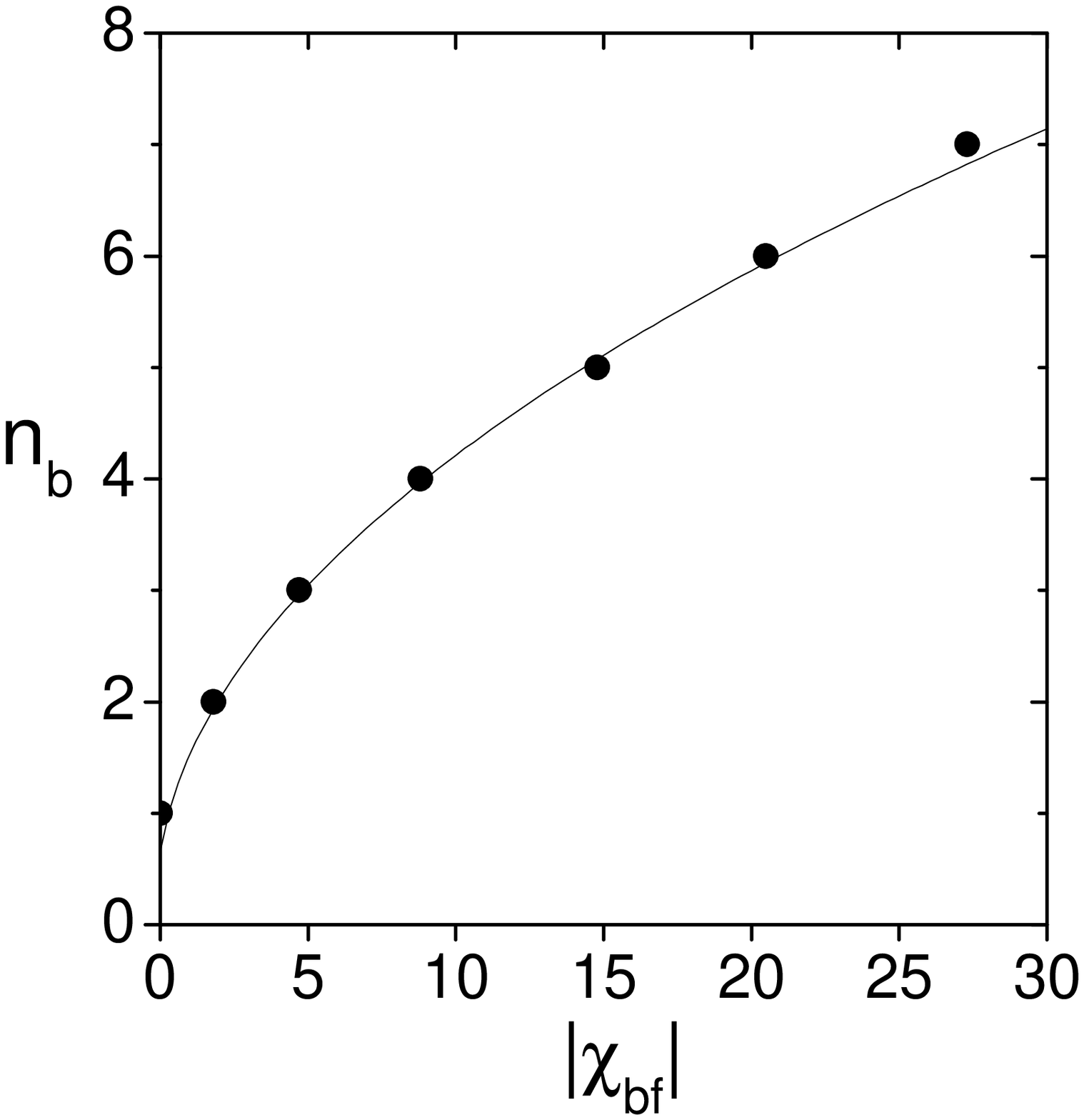}}
\caption{Left panel. Energy levels of the fermionic bound states
trapped inside an attractive condensate as a function of
$\chi_{bf}$. The continuous curves refer to numerical results
obtained with the self-consistent method while the dotted lines to
the analytical expression in Eq. (\ref{ptlevel}) with $a=0.66$ and
$n=0,1,...,6$, from right to the left, respectively. The top
horizontal line represents the bottom of the fermionic band.
Parameters are fixed as in Fig. \ref{fig7}.  Right panel. Number
of fermions entering the gap soliton  as a function of
$\chi_{bf}$. The continuous curve refers to Eq. (\ref{npt}) with
fitting parameter $a=0.66$. To account for  the existence of one
bound state at small $\chi_{bf}$ the curve has been shifted upward
by $a$. All plotted quantities are in normalized units.}
\label{fig8}
\end{figure}

Notice that with this parametrization the condition  $\int \rho_b
dx=N_b$ is always satisfied so that  $a$ can be used as a fit
parameter. In the left panel of Fig. \ref{fig7} we have compared
the bosonic density obtained using the self-consistent method with
the P\"oschl-Teller approximation  in Eq. (\ref{pt}), from which
we see that the approximation is indeed quite good. The right
panel of the figure shows the gap soliton with  four trapped
fermionic levels inside. Comparing these wavefunctions with the
ones in Fig. \ref{fig6} we see that, although different,  they
have the same symmetry properties with respect to the optical
lattice. The fact that the bosonic density can be well
approximated by a P\"oschl-Teller potential allows to solve
exactly the Schr\"odinger equation for the effective potential
$V_{eff}=-|\chi_{bf}|\rho_b$ in terms of hypergeometric functions
\cite{landau}. The vanishing of bound state wavefunctions  at
large distances requires that the hypergeometric series must be
reduced to a polynomial, this leading to the following analytical
expression for the fermionic levels inside the condensate
\begin{equation}
E_{n}=-\frac 1{4 a^2} \left(2n+1 -\sqrt{1+2 a |\chi_{bf}| N_b
}\right)^2, n=0,1,...,n_f. \label{ptlevel}
\end{equation}
The number of discrete levels $n_f$ in the gap soliton is then
obtained as the largest integer $n$ satisfying the  inequality
\begin{equation}
n < \frac 12 (\sqrt{1+2 a  |\chi_{bf}| N_b }-1). \label{npt}
\end{equation}
In Fig. \ref{fig8} we compare the energy levels obtained with the
self-consistent method with those obtained from Eq.
(\ref{ptlevel}), from which we see that the agreement is quite
good. In the right panel a similar comparison is made for the
number of fermions $n_f$ versus $\chi_{bf}$, this further
confirming the validity of our approximation. Notice that for $N_b
|\chi_{bf}|a\gg 1$ the energy levels (\ref{ptlevel}) coincide with
those obtained with the WKB  approximation valid for $n$ very
large (in this limit $n_f$ can be approximated as $n_f \approx
\sqrt{N_b |\chi_{bf}|a /2})$. From this we conclude that the
number of fermionic levels at the bottom of the Fermi sea which
can be trapped in a gap soliton can be controlled by changing the
boson-fermion scattering length using Feshbach resonances.

\section{Conclusions} In conclusion, we have shown that localized
states of BFM with attractive (repulsive) Bose-Fermi interactions
can be viewed  as a matter wave realization of quantum dots
(antidots).  The case of BFM in optical lattices has been
investigated in detail and the existence of gap solitons has been
shown.  In particular, we showed that gap-solitons can trap a
number of fermionic bound state levels inside both for repulsive
and attractive boson-boson interactions. These solutions  may have
different symmetries with respect to the OL and exist for a wide
range of parameters. Gap solitons of BFM with OS and IA type  are
very stable under time evolution. We have shown that, as the
boson-fermion interaction is changed the trapped fermionic levels
inside the condensate may undergo level crossings or avoid
crossings depending on the symmetry of the corresponding
wavefunctions with respect to the optical lattice.  This behavior
is in agreement with a general theorem of quantum mechanics and
shows that trapped fermions in BEC are in true quantum regime.
The dependence of the number of bound states and of their energies
on $N_b$ and $\chi_{bf}$ has been calculated both numerically and
analytically. In particular, for attractive boson-boson
interactions we approximated the bosonic density with a
P\"oschl-Teller potential and  showed that a gap soliton in a
shallow OL gives rise to a matter q-dot with filling number
$n_f\propto \sqrt{|\chi_{bf}|N_b}$. The possibility to change  the
boson-fermion interaction in a large interval of values (both
negative to positive) by means of a Feshbach resonance, makes the
filling dependence of a matter q-dot on $\chi_{bf}$ experimentally
feasible to check. To this regard we expect that as the (negative)
boson-fermion scattering length is decreased, the progressive
filling of the gap soliton gives rise to a depletion of fermions
in the OL which follow the above law and which could be in
principle monitored by imaging techniques. To this regard we
remark that the existence of gap-solitons in one dimensional
Bose-Einstein condensates in optical lattices has been
experimentally demonstrated in Ref.\cite{markus}. We believe that
similar experiments can be performed also for Bose-Fermi mixtures
in OL and we expect that gap solitons will be found also in this
case.

Finally, we remark that the possibility to use gap solitons of BFM
as q-bits protected from thermal dechoerence is very appealing and
deserves further investigations.

\paragraph{Acknowledgments}
Interesting discussions with F.Kh. Abdullaev and B.B. Baizakov are
gratefully acknowledged. Financial support from the
MURST-PRIN-2003 project {\it Dynamical properties of Bose-Einstein
condensates in optical lattices}  is also acknowledged.

\end{document}